\documentclass[twocolappendix,numberedappendix]{emulateapj}

\usepackage{amsmath,amssymb}

\usepackage{xspace}
\usepackage{graphics,graphicx}
\usepackage{natbib}
\usepackage{multirow}
\usepackage[percent]{overpic}
\usepackage{color}
\usepackage{bm}

\newcommand{\beq}{\begin{equation}}
\newcommand{\beqa}{\begin{eqnarray}}
\newcommand{\eeq}{\end{equation}}
\newcommand{\eeqa}{\end{eqnarray}}
\newcommand{\bfx}{\mbox{\boldmath{$x$}}}

\newcommand{\bfk}{\bm k}

\newcommand{\bfd}{\mbox{\boldmath{$d$}}}

\newcommand{\bfell}{\bm \ell}

\newcommand{\bftheta}{\bm \theta}

\shorttitle{$108$ full-sky lensing mock catalogs}
\shortauthors{Takahashi, Hamana, Shirasaki, et al.}

\begin{document}

\title{Full-sky Gravitational Lensing Simulation for Large-area Galaxy Surveys and Cosmic Microwave Background Experiments}
\author{Ryuichi Takahashi$^1$, Takashi Hamana$^2$, Masato Shirasaki$^2$, Toshiya Namikawa$^{3,4}$, Takahiro Nishimichi$^{5,6}$, \\ Ken Osato$^7$, and Kosei Shiroyama$^1$}
\affil{$^1$Faculty of Science and Technology, Hirosaki University,
  3 Bunkyo-cho, Hirosaki, Aomori 036-8561, Japan \\
  $^2$National Astronomical Observatory of Japan, Mitaka, Tokyo 181-8588, Japan \\
  $^3$Department of Physics, Stanford University, Stanford, California 94305, USA  \\
  $^4$Kavli Institute for Particle Astrophysics and Cosmology, SLAC National Accelerator Laboratory, Menlo Park, California 94025, USA
  $^5$Kavli Institute for the Physics and Mathematics of the Universe (WPI), The University of Tokyo Institutes for AdvancedStudy (UTIAS), The University of Tokyo, 5-1-5 Kashiwanoha, Kashiwa-shi, Chiba, 277-8583, Japan \\
  $^6$CREST, JST, 4-1-8 Honcho, Kawaguchi, Saitama, 332-0012, Japan \\
  $^7$Department of Physics, School of Science, The University of Tokyo, 7-3-1 Hongo, Bunkyo, Tokyo 113-0033, Japan
}

\begin{abstract}

We present 108 full-sky gravitational lensing simulation data sets generated by performing multiple-lens plane ray-tracing through high-resolution cosmological $N$-body simulations.
The data sets include full-sky convergence and shear maps from redshifts $z=0.05$ to $5.3$ at intervals of $150 \, h^{-1}{\rm Mpc}$ comoving radial distance (corresponding to a redshift interval of $\Delta z \simeq 0.05$ at the nearby universe), enabling the construction of a mock shear catalog for an arbitrary source distribution up to $z=5.3$.
The dark matter halos are identified from the same $N$-body simulations with enough mass resolution to resolve the host halos of the Sloan Digital Sky Survey (SDSS) CMASS and luminous red galaxies (LRGs).
Angular positions and redshifts of the halos are provided by a ray-tracing calculation, enabling the creation of a mock halo catalog to be used for galaxy-galaxy and cluster-galaxy lensing.
The simulation also yields maps of gravitational lensing deflections for a source redshift at the last scattering surface, and we provide 108 realizations of lensed cosmic microwave background (CMB) maps in which the post-Born corrections caused by multiple light scattering are included.
We present basic statistics of the simulation data, including the angular power spectra of cosmic shear, CMB temperature and polarization anisotropies, galaxy-galaxy lensing signals for halos, and their covariances.
The angular power spectra of the cosmic shear and CMB anisotropies agree with theoretical predictions within $5\%$ up to $\ell = 3000$ (or at an angular scale $\theta > 0.5$ arcmin).
The simulation data sets are generated primarily for the ongoing Subaru Hyper Suprime-Cam survey, but are freely available for download at http://cosmo.phys.hirosaki-u.ac.jp/takahasi/allsky\_raytracing/.

\end{abstract}


\section{Introduction}

Weak gravitational lensing is a powerful tool for probing the dark matter distribution and the nature of dark energy \citep[see, e.g., a review by][]{hj08,mvwh08,k15,b16}.
In this phenomenon, observed images of distant galaxies are coherently deformed by weak-lensing effects caused by foreground matter distribution.
These distortions enable estimation of the density fluctuations in foreground matter through the correlation of background galaxy shapes (the so-called ``cosmic shear'') and facilitate measurement of the mean mass profiles of foreground galaxies through stacking of the background shear fields (via so-called ``galaxy-galaxy lensing'').
The lensing signals are also sensitive to the amount of dark energy, in terms of both the distances to the sources (or lenses) and the growth rate of the density fluctuations.

The Canada-France-Hawaii Telescope Lensing Survey\footnote{ http://www.cfhtlens.org} (CFHTLens) is a state-of-the-art weak-lensing survey that has observed $150 \, {\rm deg}^2$ and provided strong constraints on the amplitude of density fluctuations and the matter density parameter \citep[e.g.,][]{hey12,hey13,k13}.
Several other larger-area surveys are also currently in progress.
In the northern sky, the Subaru Hyper Suprime-Cam survey\footnote{http://www.naoj.org/Projects/HSC/} (the HSC survey, hereafter) started their survey of $1400 \, {\rm deg}^2$ in 2014 \citep{m12,hsc17}, and in the southern sky, the Dark Energy Survey\footnote{https://www.darkenergysurvey.org/} (DES) and the Kilo-Degree Survey\footnote{http://kids.strw.leidenuniv.nl/} (KiDS) are going to cover $5000$ and $1500 \, {\rm deg}^2$, respectively \citep{dej15,des16a}.
DES and KiDS have already presented their first science results of cosmological parameter constraints obtained from cosmic shear data \citep{des16b,hild17}.
Beginning in 2020, the Large Synoptic Survey Telescope\footnote{https://www.lsst.org/} (LSST) plans to image half of the southern sky \citep[e.g.,][]{cha13}.
In the 2020s, { the} space mission Euclid\footnote{http://sci.esa.int/euclid/} and the Wide Field Infrared Survey Telescope\footnote{https://wfirst.gsfc.nasa.gov/} (WFIRST) plan to start large galaxy surveys.

The foreground matter also deforms the temperature and polarization patterns on the sky of the cosmic microwave background (CMB).
The lensing of the CMB violates parity symmetry and produces a rotational pattern (B-mode) in CMB polarization \citep{zs98}. 
The CMB lensing contains information on density fluctuations at relatively high redshift $z \simeq 1-3$ \citep[e.g.,][]{lc06}.
Precise measurements and removal of lensing-induced B-modes will be required to enhance sensitivity to B-modes that are generated by primordial gravitational waves \citep[e.g.,][]{Kesden02,Knox02}. 
The lensing signature was first detected by cross-correlating the Wilkinson Microwave Anisotropy Probe (WMAP) temperature map with a foreground galaxy distribution \citep{szd07,h08}.
Lensing signals have also been detected using CMB data alone from multiple CMB experiments \citep{pb14,spt14,bk8,planck16xv,act17}.
In ongoing and near-future CMB experiments such as those reported by the Simons Observatory\footnote{https://simonsobservatory.org/}, SPT-3G\footnote{https://pole.uchicago.edu/spt/}, and CMB-S4\footnote{https://cmb-s4.org/}, the nonlinear growth of the large-scale structure and post-Born corrections become important in the lensing analysis \citep{boehm16,pl16} and are detectable by measuring the bispectrum \citep{nami16} and higher-order cumulants \citep{Liu16}. 
As these experiments will observe a wide area of CMB sky, a full-sky simulation including the effect of the nonlinear evolution will be required. 

To extract cosmological information from the observational data, a large number of catalogs are necessary to estimate the covariances of observables \citep[e.g.,][]{taka09}.
Mocks are useful in testing analysis tools that are to be used in forthcoming real observations.
Because current weak-lensing surveys and CMB experiments cover large areas ($> 1000 \, {\rm deg}^2$), full-sky mocks covering the entire survey regions will be required.
Furthermore, CMB and galaxy shear maps on the same sky region are not independent because they trace the same foreground matter distribution; therefore, mock catalogs for different probes that are consistently constructed out of the same foreground density fields are necessary for cross-correlation analyses.
The observables and their covariances generally depend on the survey geometry and mask regions in a complicated manner \citep[e.g.,][]{th13};
as such, full-sky mocks will be useful for correctly incorporating this geometrical dependence into the analysis.

As weak lensing probes small scales ($\lesssim 10 \, {\rm Mpc}$), it is necessary to apply an $N$-body simulation-based ray-tracing technique to investigate the nonlinear gravitational evolution \citep[e.g.,][]{jsw00,hm01,vw03,hilb09}.
Through such ray-tracing simulations, non-Gaussian error can be naturally introduced from mode coupling between different scales into the cosmic-shear power spectrum \citep{ch01}.
Furthermore, it is possible in such simulations to identify individual halos, which is necessary to locate the formation sites of galaxies for the study of galaxy-galaxy lensing.
Previous authors have prepared a large number ($\gtrsim 100$) of mock catalogs to estimate the covariances of observables such as the convergence power spectrum \citep{sato09,kies11,hvv12,hv15,p16b}, the correlation function \citep{sato11,hvv12,hv15}, galaxy-galaxy lensing \citep{shy15,stm16}, and the three-point function (or bispectrum) of the convergence \citep{kayo13,sn13}.
These mocks have also been used to estimate the probability distributions of the convergence or magnification \citep[e.g.,][]{hmf00,tth02,do06,hilb07,taka11} and the higher-order moments (skewness and kurtosis) of the convergence \citep[e.g.,][]{p16b,sm17}.

Full-sky lensing simulations have previously been performed in CMB lensing studies \citep{car08,car09,db08,seh10,vane14} and in studies of the statistical properties of the convergence field \citep{f08,tey09,b13,f15a,f15b,h15,shy15}.
The full-sky simulation we present here has several advantages over previous works:
(1) because we generated a set of simulation data (weak-lensing maps, lensed CMB maps, and halo catalogs) from the same underlying density field, it is possible to study their cross-correlations in the form of galaxy-galaxy lensing, CMB lensing with cosmic shear, and CMB lensing with halo clustering; (2) because we do not employ the Born approximation, it is possible to study the post-Born effects in CMB lensing (note that very recently, \cite{fcc17} also studied this topic); (3) we produced the largest sample of full-sky lensing simulation data to date, with $108$ generated realizations. 

Our simulation is designed primarily for the preparation of weak-lensing mock catalogs of the HSC survey.
Because this survey covers $\sim 1400$ sq.~degrees, $20$ mock HSC survey regions can be taken from a single full-sky map without overlap \citep{shy15}, translating into $\sim 2000$ HSC mocks in total from the $108$ full-sky maps.
This large number enables us to calculate the covariances of observables with a reasonable accuracy ($\sim (2000/2)^{-1/2} \simeq 3 \%$, see also Section 4).
The simulation also resolves dark matter halos whose mass range comprehensively covers the SDSS CMASS galaxies of typical halo mass $\sim 10^{13} h^{-1} {\rm M}_\odot$ at $0.4<z<0.8$ \citep[e.g.,][]{w11,a15} and luminous red galaxies (LRGs) of halo mass $\sim 10^{13-14} h^{-1} {\rm M}_\odot$ at $z<0.5$ \citep[e.g.,][]{e01,zh09}, making it possible to study galaxy-galaxy lensing and the cross correlation of the halo distribution with the CMB map.

In Section 2 we describe in detail our numerical simulations and simulation products (which are further described in the appendices).
We present the basic statistics of our data and compare them with theoretical models in Sections 3 and 4.
In Section 5 we discuss known issues of our simulation data.
A summary and discussion are presented in Section 6, and a guide to use the simulation products is provided in Appendix D and on the web\footnote{http://cosmo.phys.hirosaki-u.ac.jp/takahasi/allsky\_raytracing/}.
Throughout this paper, we adopt the standard $\Lambda$CDM (Lambda cold dark matter) cosmology  that is consistent with the WMAP $9$ years result \citep{h13}.
The cosmological parameters are the CDM density parameter $\Omega_{\rm cdm}=0.233$, the baryon density $\Omega_{\rm b}=0.046$, the matter density $\Omega_{\rm m}=\Omega_{\rm cdm}+\Omega_{\rm b}=0.279$, the cosmological constant $\Omega_\Lambda=0.721$, the Hubble parameter $h=0.7$, the amplitude of density fluctuations $\sigma_8=0.82$, and the spectral index $n_{\rm s}=0.97$.
We also adopt the natural units $c=G=1$.


\section{All-sky Ray-tracing simulation}

\begin{figure}
\epsscale{1.1}
\plotone{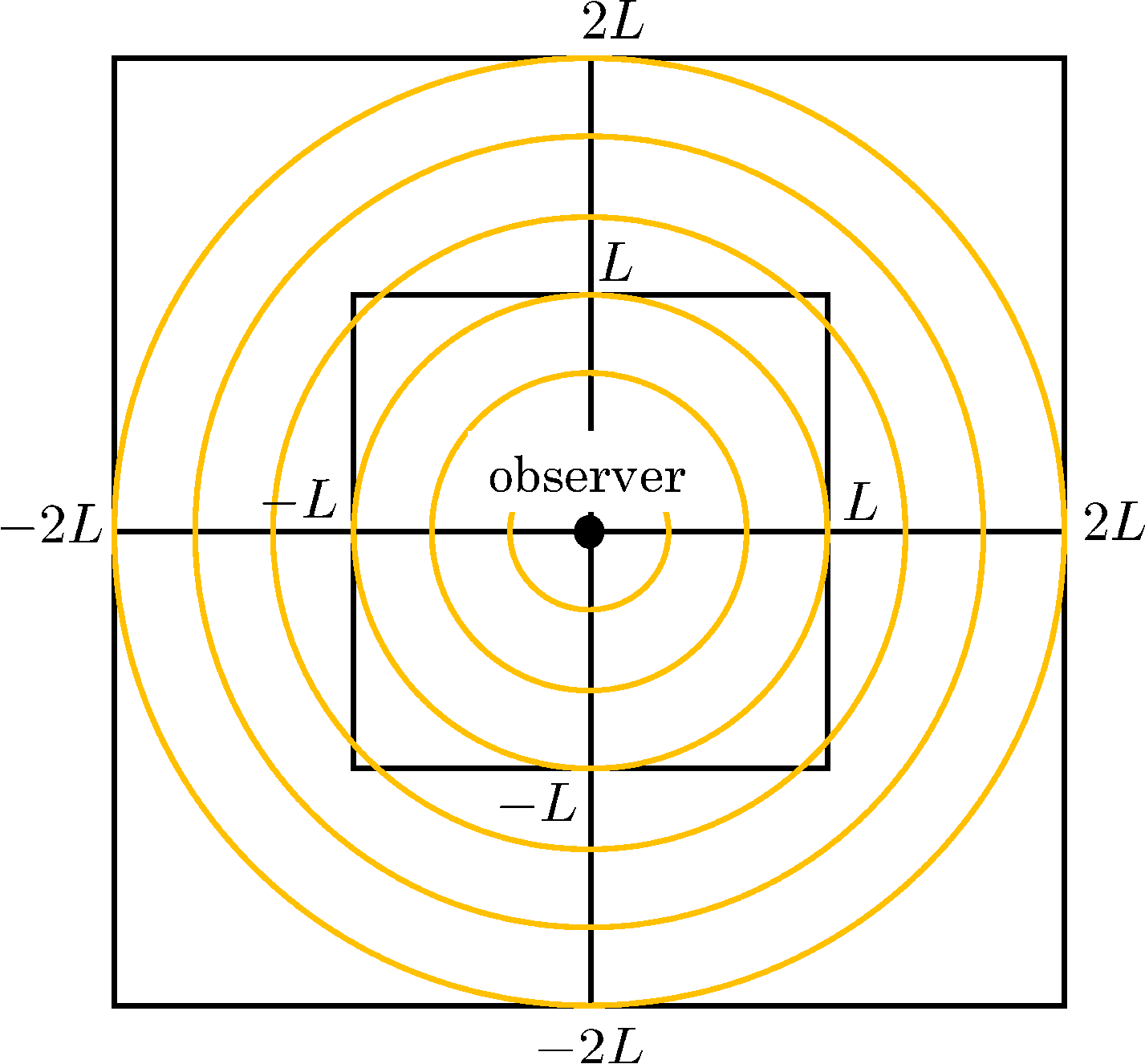}  
\caption{Configuration of ray-tracing simulation with cubic simulation boxes of lengths $L,2L,3L,\cdots$, where $L=450 \, h^{-1}$Mpc (comoving scale), placed around the observer.
  The figure shows the inner two boxes with side lengths $L$ and $2L$, respectively.
  The observer is located at the vertex of the boxes.
  In each box, we constructed three spherical shells with thickness of $\Delta r = L/3 = 150 \, h^{-1}$Mpc; the orange circles show the boundaries between the shells.}
\label{fig_shell}
\vspace*{0.5cm}
\end{figure}

In this section, we describe our ray-tracing simulation.
We prepared a system of nested cubic simulation boxes to reproduce the mass distribution in the Universe.
The boxes had side lengths $L,2L,3L,\cdots$, with $L=450 \, h^{-1}$Mpc, and were placed around a fixed vertex representing the observer's position, as shown in Figure \ref{fig_shell}.
In this scheme, each box was duplicated eight times and placed around the observer using the periodic boundary conditions, and we constructed three spherical lens shells of width $\Delta r = 150 \, h^{-1}$Mpc within each box period (see Figure \ref{fig_shell}).
We then calculated the gravitational potential on the lens shells and traced the resulting light-ray paths from the observer to the last scattering surface.
For basic equations and details of the multiple-plane gravitational lensing algorithm and actual implementations for the numerical simulation, we refer the reader to Appendix C of \cite{shy15}.

\subsection{N-body Simulations}

\begin{deluxetable*}{cccccc}
\tablecaption{$N$-body Simulation Parameters}
\startdata
\hline 
 Box Size & Particle Mass & Minimum Halo Mass & Softening Length  & Dumped & Initial  \\
  ($h^{-1}$Mpc) & ($h^{-1} {\rm M}_\odot$) & ($h^{-1} {\rm M}_\odot$) & ($h^{-1} {\rm kpc}$) & Redshift & Redshift  \\
  \hline
 $450$   & $8.2 \times 10^8$  & $4.1 \times 10^{10}$  & $8$ &  $0.025,0.076,0.129$  & $63$  \\
 $900$   & $6.6 \times 10^9$  & $3.3 \times 10^{11}$  & $16$ &  $0.182,0.238,0.294$ & $63$  \\
 $1350$  & $2.2 \times 10^{10}$ & $1.1 \times 10^{12}$ & $24$ & $0.353,0.413,0.476$  & $63$  \\
 $1800$  & $5.3 \times 10^{10}$ & $2.6 \times 10^{12}$ & $32$ & $0.541,0.608,0.678$  & $49$  \\
 $2250$  & $1.0 \times 10^{11}$ & $5.1 \times 10^{12}$ & $40$ & $0.751,0.827,0.907$  & $49$  \\
 $2700$  & $1.8 \times 10^{11}$ & $8.9 \times 10^{12}$ & $48$ & $0.990,1.078,1.170$  & $49$  \\
 $3150$  & $2.8 \times 10^{11}$ & $1.4 \times 10^{13}$ & $56$ & $1.267,1.370,1.478$  & $39$  \\
 $3600$  & $4.2 \times 10^{11}$ & $2.1 \times 10^{13}$ & $64$ & $1.593,1.714,1.844$  & $39$  \\
 $4050$  & $6.0 \times 10^{11}$ & $3.0 \times 10^{13}$ & $72$ & $1.982,2.130,2.287$  & $39$  \\
 $4500$  & $8.2 \times 10^{11}$ & $4.1 \times 10^{13}$ & $80$ & $2.456,2.638,2.833$  & $31$  \\
 $4950$  & $1.1 \times 10^{12}$ & $5.5 \times 10^{13}$ & $88$ & $3.044,3.272,3.519$  & $31$  \\
 $5400$  & $1.4 \times 10^{12}$ & $7.1 \times 10^{13}$ & $92$ & $3.788,4.080,4.400$  & $31$  \\
 $5850$  & $1.8 \times 10^{12}$ & $9.0 \times 10^{13}$ & $100$ & $4.750,5.135,5.560$  & $31$  \\
 $6300$  & $2.3 \times 10^{12}$ & $1.1 \times 10^{14}$ & $108$ & $6.030,6.551,7.133$  & $31$ 
  \enddata
\tablecomments{
 The side length of the simulation box, the particle mass, the minimum halo mass, the softening length, the dumped redshift, and the initial redshift. { The number of particles and realizations are $2048^3$ and $6$, respectively, for all the simulation boxes.}
}
\label{nbody}
\end{deluxetable*}

We first conducted a cosmological $N$-body simulation on the periodic cubic box following the gravitational evolution of dark matter particles without baryonic processes.
We generated the initial conditions based on the second-order Lagrangian perturbation theory \citep[2LPT;][]{cps06,n09} with the initial linear power spectrum calculated using the Code for Anisotropies in the Microwave Background \citep[CAMB;][]{lcl00}.
We followed the gravitational evolution from the initial redshift using the $N$-body code \textsc{Gadget2} \citep{syw01,s05} with a softening length set to $4\%$ of the mean particle separation.
We prepared $14$ boxes with side lengths $L=450$, $900$, $1350$, $\cdots$, $6300 \, h^{-1}$Mpc in steps of $450 \, h^{-1}$Mpc, with six independent copies at each box size, totaling $84 \, (=14 \times 6)$ realizations in total.
The number of particles for each box was $2048^3$, making the mass and spatial resolutions better for the inner boxes.
The particle positions were dumped at redshifts corresponding to the comoving distances to the lens planes at $r=150 \, (i-0.5) \, h^{-1} {\rm Mpc}$ for $i=1,2,3, \cdots$.
Table \ref{nbody} summarizes our simulation settings, including box side lengths, particle masses, minimum halo masses (which were $50$ times the particle masses), softening lengths, dumped redshifts, and initial redshifts.

We checked that the average matter power spectra of all six realizations agreed with  theoretical predictions of the revised Halofit \citep{s03,t12} within $5\% \, (10\%)$ for $k<5 \, (6) \, h{\rm Mpc}^{-1}$ at $z<1$, for $k<0.8 \, (1) \, h{\rm Mpc}^{-1}$ at $z<3$, and for $k<0.5 \, (0.7) \, h{\rm Mpc}^{-1}$ at $z<7$. 
Figure \ref{fig_pk} in Appendix A shows a comparison of the measured power spectra with the Halofit prediction. The figure clearly shows the scales that are resolved.

\subsection{Ray-tracing Simulation}

\begin{figure}
\hspace*{-0.7cm}
\includegraphics[width=1.1\columnwidth]{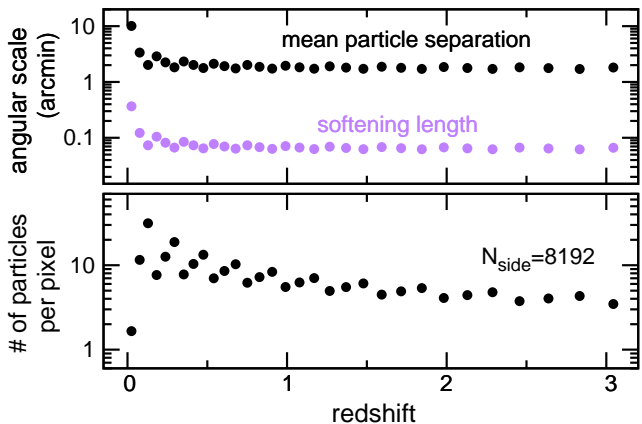} 
\vspace*{-0.7cm}
\caption{
  {\bf Upper panel:} angular scales of mean particle separation (black dots) and softening length (purple dots) projected onto the lens plane.
  The x-axis is the redshift of the lens plane from $z=0$ to $3$.
  The softening length is $4 \%$ of the mean particle separation.
  {\bf Lower panel:} particle number density per pixel on the lens shell at resolution $N_{\rm side}=8192$.  
  The number density simply scales as $\propto N_{\rm side}^{-2}$.
}
\label{fig_theta_Npart}
\end{figure}

We briefly explain the procedure we used to trace light rays through the $N$-body data.
We used the public code \textsc{GRayTrix},\footnote{http://th.nao.ac.jp/MEMBER/hamanatk/GRayTrix/} which follows the standard multiple-lens plane algorithm in spherical coordinates (see Appendix C of \cite{shy15} for a detailed description of the algorithm) using the \texttt{HEALPix} algorithm \citep{g05}.
As shown in Figure \ref{fig_shell}, we constructed three lens shells in each simulation box and projected the particle positions onto these shells.
There were $12 \times N_{\rm side}^2$ equal-area pixels on each shell in the \texttt{HEALPix} pixelization, and we computed the projected surface density by assigning each particle to the nearest pixel.
The two-dimensional gravitational potential was solved via the Poisson equation using the multipole expansion.
The deflection angle and Jacobian matrix were obtained from the first and second covariant derivatives of the potential \citep{b13} using the \texttt{HEALPix} subroutines \texttt{map2alm} and \texttt{alm2map$\_$der} to perform the computation.
In our simulation, we adopted the three-pixel resolutions $N_{\rm side}=4096$, $8192$, and $16384$, corresponding to pixel sizes of $0.86$, $0.43$, and $0.21$ arcmin, respectively.
We chose $18$ observer's positions in each simulation box to increase the number of realizations\footnote{The $18$ positions are $(x,y,z)$ $=(0,0,0)$, $(L/2,0,0)$, $(0,L/2,0)$, $(0,0,L/2)$, $(L/2,L/2,0)$, $(0,L/2,L/2)$, $(L/2,0,L/2)$, $(L/2,L/2,L/2)$, $(\pm L/4,0,0)$, $(0,\pm L/4,0)$, $(0,0,\pm L/4)$, $(L/4,L/4,0)$, $(0,L/4,L/4)$, $(L/4,0,L/4)$, and $(L/4,L/4,L/4)$.} (it is possible to choose an arbitrary observer's position within the box under the periodic boundary condition), resulting in $18 \times 6 = 108$ realizations of lens shells at each redshift. 
We prepared $108$ lens shells for $N_{\rm side}=4096$ and $8192$ and a single shell for $N_{\rm side}=16384$.
The simulation boxes were prepared up to redshift $7.1$ (see Table \ref{nbody}); for higher redshifts up to the last scattering surface ($z = 7.1-1100$), we constructed lens shells assuming Gaussian surface density fluctuations based on the linear matter power spectrum.

The upper panel of Figure \ref{fig_theta_Npart} plots the angular scales of mean particle separation (black dots) and softening length (purple dots) in the $N$-body data against the redshift of the lens plane ($x$-axis).
As the angular scale of the mean particle separation (softening length) is typically $\sim 2 \, (0.08)$ arcmin for $z \gtrsim 0.1$, the softening length is well below the angular pixel size.
The lower panel plots the number of particles per pixel on the lens shell; the typical number density is about $2-10$ for $N_{\rm side}=8192$ at $z \gtrsim 0.3$.
Here the number density simply scales as $\propto N_{\rm side}^{-2}$ for other pixel resolutions. 

Finally, we traced the light-ray paths by solving lens equations to obtain the angular positions of light rays on the source planes.
We also obtained full-sky maps of the convergence $\kappa$, shear $\gamma_{1,2}$, and rotation $\omega$ by solving the evolution equation of the Jacobian matrix along the light-ray path.
Here, the Jacobian matrix is 
  \beq
  A = \left(
    \begin{array}{cc}
      1-\kappa-\gamma_1 & -\gamma_2-\omega \\
      -\gamma_2+\omega & 1-\kappa+\gamma_1
    \end{array}
    \right).
  \label{jacob_mat}  
  \eeq
The lens planes were placed at $r=150 \, (i-0.5) \, h^{-1} {\rm Mpc}$, while the source planes were at $r=150 \, i \, h^{-1} {\rm Mpc}$, where $i=1,2,\cdots$.
These full-sky maps were prepared at every $150 \, h^{-1}$Mpc comoving distance up to $z=5.3$ and at the last scattering surface ($z=1100$).
Table \ref{source_redshift} shows the output ID number of the source plane, the source redshift, and the comoving distance to the source.
As mentioned above, we prepared $108$ independent maps up to $z=1100$ for $N_{\rm side}=4096$ and $8192$ and a single map up to $z=5.3$ for $N_{\rm side}=16384$.
The high-resolution map with $N_{\rm side}=16384$ will be used to check the numerical convergence at small scale (near the pixel size).

Note that we randomly chose the lens planes to perform the ray-tracing simulation but avoided using the same plane more than once.
As we used three lens planes taken from each given box as a single set, the matter distribution along the line-of-sight direction is continuous at the boundaries between the lens planes within each box, but discontinuous between planes taken from different-sized boxes.
As a result, the radial structure appears continuous within the effective shell thickness of $450 \, h^{-1}$Mpc (not $150 \, h^{-1}$Mpc).
Note that the thickness of the lens shell slightly affects the convergence and shear power spectra (as we discuss in Section 3 and in Appendix B).


\begin{deluxetable}{ccc}
\startdata
\hline 
 output & source  &  distance    \\
 ID number &  redshift &  $(h^{-1}{\rm Mpc})$  \\
  \hline
   zs1    &    $0.051$    &     $150$  \\
   zs2    &    $0.102$    &     $300$  \\
   zs3    &    $0.155$    &     $450$  \\
   zs4    &    $0.210$    &     $600$  \\
   zs5    &    $0.266$    &     $750$  \\
   zs6    &    $0.323$    &     $900$  \\
   zs7    &    $0.383$    &    $1050$  \\
   zs8    &    $0.444$    &    $1200$  \\
   zs9    &    $0.508$    &    $1350$  \\
   zs10   &    $0.574$    &    $1500$  \\
   zs11   &    $0.643$    &    $1650$  \\
   zs12   &    $0.714$    &    $1800$  \\
   zs13   &    $0.788$    &    $1950$  \\
   zs14   &    $0.866$    &    $2100$  \\
   zs15   &    $0.948$    &    $2250$  \\
   zs16   &    $1.033$    &    $2400$  \\
   zs17   &    $1.123$    &    $2550$  \\
   zs18   &    $1.218$    &    $2700$  \\
   zs19   &    $1.318$    &    $2850$  \\
   zs20   &    $1.423$    &    $3000$  \\
   zs21   &    $1.535$    &    $3150$  \\
   zs22   &    $1.653$    &    $3300$  \\
   zs23   &    $1.778$    &    $3450$  \\
   zs24   &    $1.912$    &    $3600$  \\
   zs25   &    $2.055$    &    $3750$  \\
   zs26   &    $2.207$    &    $3900$  \\
   zs27   &    $2.370$    &    $4050$  \\
   zs28   &    $2.546$    &    $4200$  \\
   zs29   &    $2.734$    &    $4350$  \\
   zs30   &    $2.937$    &    $4500$  \\
   zs31   &    $3.156$    &    $4650$  \\
   zs32   &    $3.393$    &    $4800$  \\
   zs33   &    $3.651$    &    $4950$  \\
   zs34   &    $3.931$    &    $5100$  \\
   zs35   &    $4.237$    &    $5250$  \\
   zs36   &    $4.571$    &    $5400$  \\
   zs37   &    $4.938$    &    $5550$  \\
   zs38   &    $5.342$    &    $5700$  \\
  -----   &    -----      &   -----    \\
   zs66   &    $1100$     &    $9900$    
\enddata
\tablecomments{
  The output ID number of the source plane, the source redshift, and the comoving distance to the source. There are $108$ full-sky maps each for $N_{\rm side}=4096$ and $8192$, and there is a single map for $N_{\rm side}=16384$ from zs1 to zs38.
}
\label{source_redshift}
\end{deluxetable}

\subsection{Halo Catalogs}

\begin{figure}
\epsscale{1.2}
\plotone{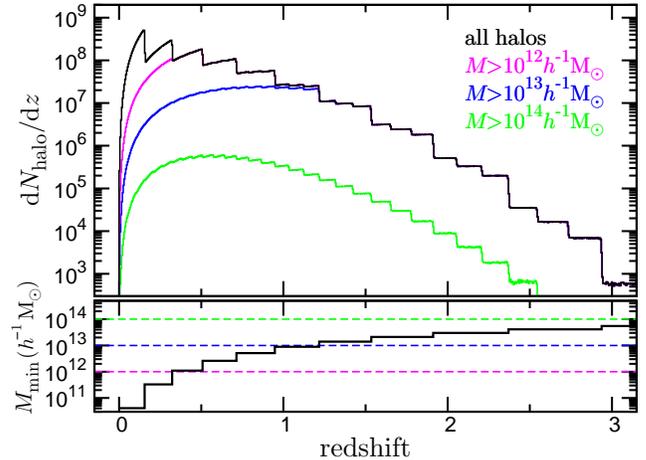}  
\caption{{\bf Upper panel:} halo number distribution as a function of redshift.
  The vertical axis shows the halo number at redshift interval $z$ to $z+dz$ in the overall sky.
  The black line is the result for all the halos, while the red, blue, and green lines are for heavier halos $M_{200{\rm b}}>10^{12} \, h^{-1}{\rm M}_\odot$, $10^{13} \, h^{-1}{\rm M}_\odot$, and $10^{14} \, h^{-1}{\rm M}_\odot$, respectively.
  {\bf Lower panel:} minimum halo mass ($M_{200 {\rm b},{\rm min}}$) as a function of redshift. The horizontal dashed lines denote $M_{200 {\rm b}}=10^{12} \, h^{-1}{\rm M}_\odot$ (magenta), $10^{13} \, h^{-1}{\rm M}_\odot$ (blue), and $10^{14} \, h^{-1}{\rm M}_\odot$ (green), respectively.
 }
\label{fig_dndz_halo}
\vspace*{0.3cm}
\end{figure}

\begin{figure}
  \begin{minipage}{\columnwidth}
  \begin{center}
    \includegraphics[width=\columnwidth]{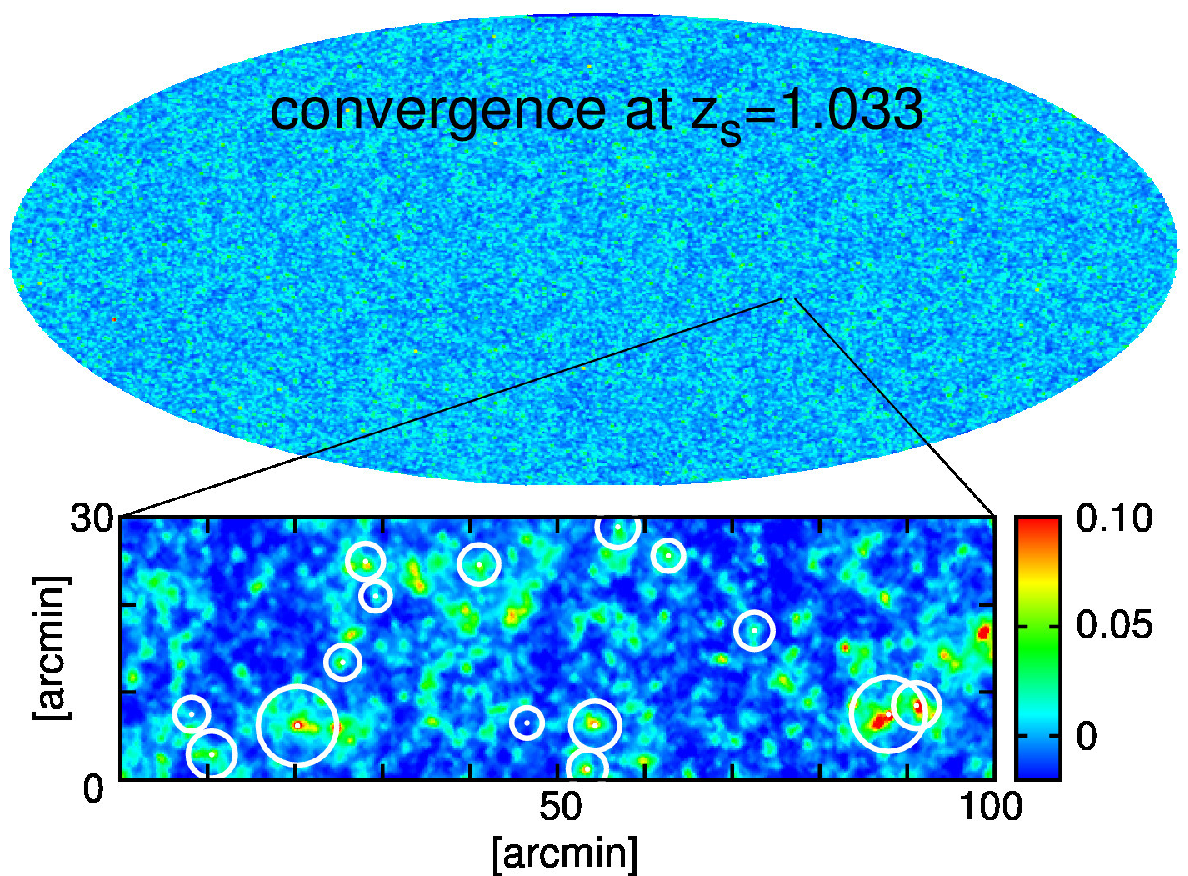}  
  \end{center}
  \vspace*{0.1cm}
  \end{minipage}

  \begin{minipage}{\columnwidth}
  \begin{center}
    \includegraphics[width=\columnwidth]{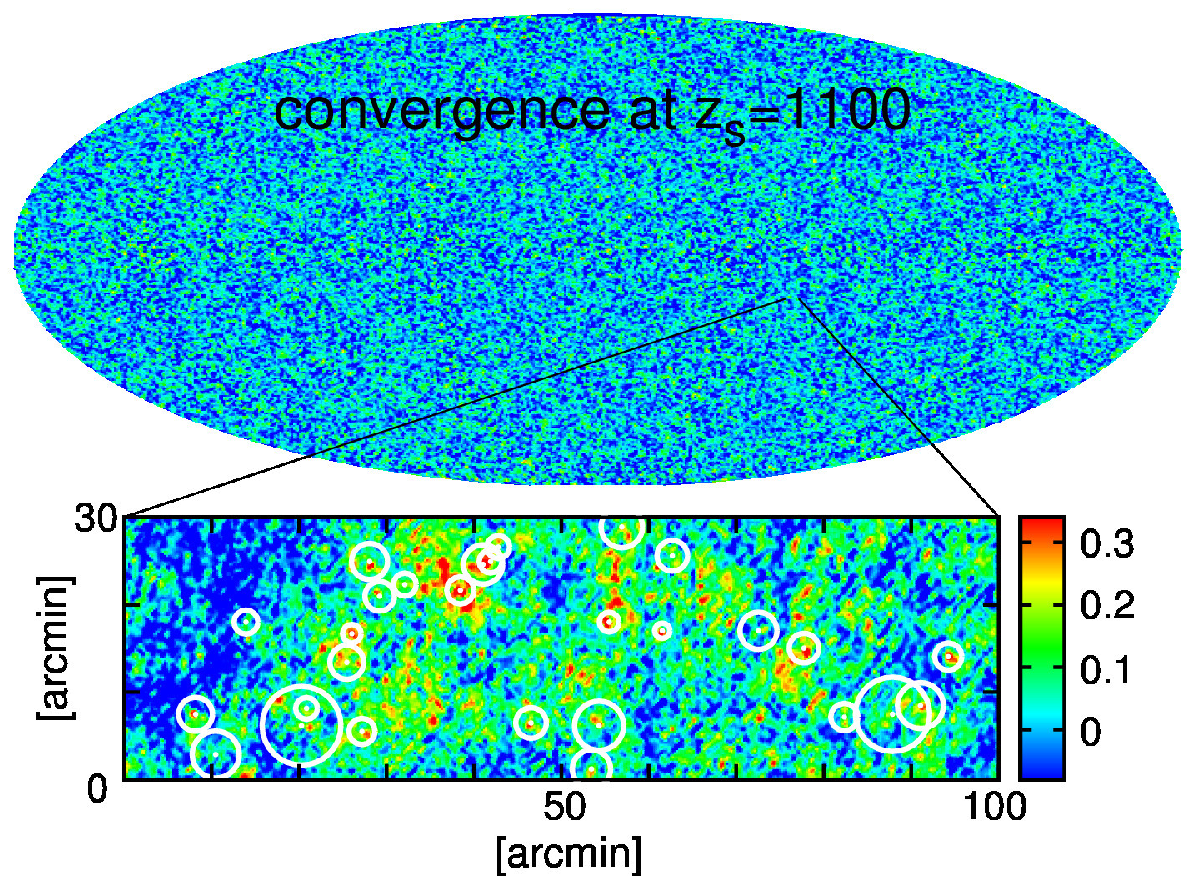}  
  \end{center}
  \vspace*{0.1cm}
  \end{minipage}
  
  \begin{minipage}{\columnwidth}
  \begin{center}
    \includegraphics[width=\columnwidth]{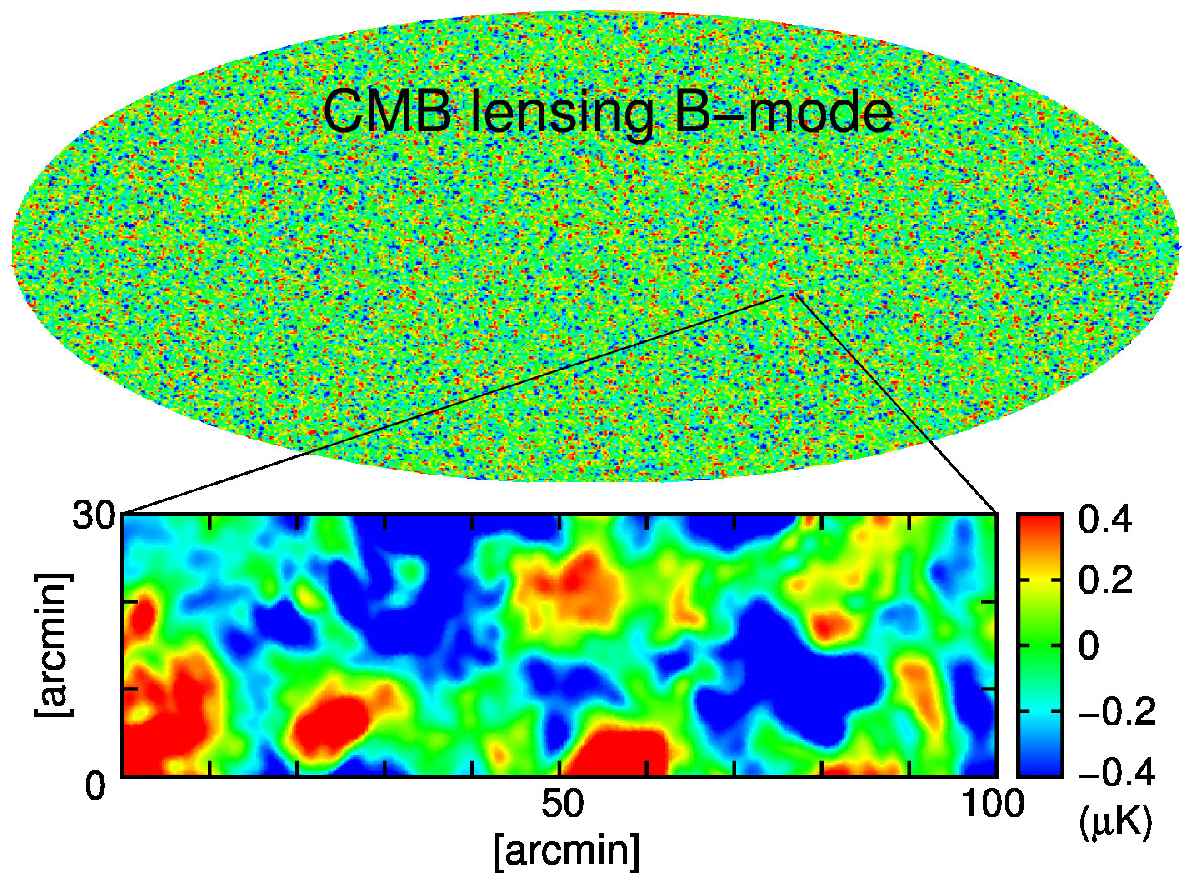}  
  \end{center}
  \vspace*{0.1cm}
  \end{minipage}
\caption{
 {\bf Top panel}: contour plot of the convergence at source redshift $z_{\rm s}=1.033$. The lower rectangular panel is a zoom-in map of size $100 \times 30 \, {\rm arcmin}^2$. The white dots with circles indicate positions of foreground massive halos of $M_{\rm 200b}>6 \times 10^{13} h^{-1} {\rm M}_\odot$, where the circle radius corresponds to the virial radius on the sky.
 {\bf Middle panel}: same as the top panel, but at the last scattering surface ($z_{\rm s}=1100$).
 {\bf Bottom panel}: same as the above panels, but for the CMB lensing B-mode. The three full-sky maps in the figure are taken from the same realization, and the three rectangular panels also show the same region. 
}
\label{fig_sample_maps}
\end{figure}

We constructed halo catalogs using the public code \textsc{Rockstar} \citep{bww13}, which identifies halos based on the six-dimensional phase-space friends-of-friends algorithm.
The masses of the identified structures were then determined according to the spherical overdensity around the center.
The minimum halo mass, given in Table \ref{nbody}, was set to $50$ times the particle mass.
The catalogs effectively cover the SDSS CMASS and LOWZ galaxy ranges, in which the halo masses are $M \gtrsim 10^{13} {\rm M}_\odot$.
Smaller halos can be resolved at lower redshifts.
The following halo data provided by \textsc{Rockstar}'s default outputs were stored: the virial mass $M_{\rm vir}$; the spherical overdensity masses within several radii, $M_{200 {\rm b}}$ (the radius at a mean halo density $200$ times larger than the background density) and $M_{200{\rm c},500{\rm c},2500{\rm c}}$ (the radii at $200,500$, and $2500$ times larger than the cosmological critical density, respectively); the halo central position and velocity; the virial radius and scale radius obtained by fitting the Navarro-Frenk-White (NFW) profile \citep{nfw97}; and the halo ID. 
The catalogs also contain subhalos and their parent halo IDs (we used the \textsc{Rockstar} utility \texttt{find\_parents} to identify these subhalos).

The \textsc{Rockstar} halo catalogs of each cubic box were combined in layers of lens shells to form full-sky light-cone halo catalogs.
To convert the (three-dimensional) halo positions and velocities into projected angular positions and radial velocities on the sky, we used the light-ray paths obtained from the ray-tracing simulation (see Appendix C3 of \cite{shy15} for technical details of the conversion).
The redshifts of halos were derived from the radial distances and peculiar velocities. 
The angular halo positions are given both on the image plane (i.e., the first lens plane) and the lens plane onto which the halo is projected.
As we used $18$ spherical shells from each simulation box, a given halo appears several times on the various maps; however, as the halos are projected onto different directions, they have different orientations.

We checked that the halo mass function in our simulation agrees with the Tinker mass function \citep{t08} within $12\%$ for $z<1.2$ and $30\%$ for $z<2.0$.
We also checked that the linear halo bias measured in our simulation is consistent with the fitting formula \citep{t10} within $10 \%$ at $k<0.4 \, h {\rm Mpc}^{-1}$ and $z<0.5$ for $M_{200 {\rm b}}=10^{12} \, h^{-1} {\rm M}_\odot$, at $k<0.4 \, h {\rm Mpc}^{-1}$ and $z<1.4$ for $M_{200 {\rm b}}=10^{13} \, h^{-1} {\rm M}_\odot$, and at $k<0.3 \, (0.1) \, h {\rm Mpc}^{-1}$ and $z<0.9 \, (1.5)$ for $M_{200 {\rm b}}=10^{14} \, h^{-1} {\rm M}_\odot$, respectively.
Figures \ref{fig_massf} and \ref{fig_bias} in Appendix A plot comparisons of the simulation results with the theoretical fitting formulae for the halo mass function and the halo bias, respectively. These figures clearly show the halo-mass and redshift ranges that are validated.

The upper panel of Figure \ref{fig_dndz_halo} plots the halo number density on the full sky as a function of redshift, with the y-axis showing the number of halos in the redshift interval $z$ to $z+dz$.
The figure shows the average number density of the $108$ maps.
The black line represents all halos heavier than the minimum halo mass, while the magenta, blue, and green lines indicate halos heavier than $10^{12}$, $10^{13}$, and $10^{14} \, h^{-1}{\rm M}_\odot$, respectively.
Thus, our simulation can resolve halos of mass $10^{12}$, $10^{13}$, and $10^{14} \, h^{-1} {\rm M}_\odot$ up to $z=0.3$, $1.2$, and $3.5$, respectively (see also the lower panel, which plots the minimum halo mass as a function of redshift).
The total halo number is $\sim 1.3 \times 10^8$ up to $z=6$ in a single map.
The halo number density is discontinuous at several redshifts because (i) the dumped redshifts of the $N$-body simulation are discrete, and (ii) the minimum halo mass suddenly changes owing to changes in the mass resolution (see Table \ref{nbody}).

The top and middle panels of Figure \ref{fig_sample_maps} show contour plots of the convergence at $z_{\rm s}=1.033$ and $1100$, respectively.
Each panel shows a full-sky map with a zoom-in map ($100 \times 30$ ${\rm arcmin}^2$).
The resolution of the map shown here is $N_{\rm  side}=8192$.
In the zoom-in maps, the white dots with circles indicate the foreground halo positions, with the radius of each white circle corresponding to the virial radius of the corresponding halo on the sky.
As clearly seen in the figure, the massive halos correspond to the high convergence peaks (we will calculate the mean convergence profiles of the halos in Section 3.3).
However, some halos do not correspond to any high peak because they are very close to the observer or to the source and therefore the lensing efficiency is low.
Small-scale noisy fluctuations on the zoom-in map can be seen at $z_{\rm s}=1100$ as a result of shot noise caused by the finite number density of dark matter particles in the $N$-body simulation.
As the number density of particles decreases for higher redshifts (see Table \ref{nbody}), the shot noise appears more significant for higher source redshifts.

\subsection{Lensed CMB Maps}

In this subsection, we present the procedure we used to construct the lensed CMB temperature and polarization maps.
The procedure is as follows: (1) we prepared the unlensed CMB temperature and polarization spectra using CAMB, ignoring any primordial B-modes; (2) based on the input angular power spectrum, we constructed $108$ unlensed maps with Gaussian fluctuations using the \texttt{synfast} routine in \texttt{HEALPix} with $N_{\rm side}=8192$; and (3) finally, we evaluated the angular position shifts on the last scattering surface using the ray-tracing simulations with $N_{\rm side}=4096$ and $8192$.
Under the transformations in step (3), the temperature and polarization fields were moved from angular position $\bftheta$ to $\bftheta+\bfd$, where $\bfd$ is the deflection angle.
To do this, we required accurate interpolation on the pixels (because the lensed ray positions were generally not at the centers of pixels) and also had to take the rotation of the polarization basis into account, which was dependent on the position on the sphere (this rotation was significant around the north and south poles).
To do this, we employed the \texttt{LensPix} scheme \citep{cl05} to transform the maps\footnote{We used the subroutine \texttt{HealpixInterpLensedMap\_GradPhi} in \texttt{LensPix} with the parameters $\ell_{\rm max}=3N_{\rm side}$ and interp\_factor $=1.5$. We slightly modified the code to read the deflection-angle data obtained in our simulation.}.
Furthermore, we used the multiple-lens scattering (instead of the Born approximation), which generated a non-zero rotation of polarization \citep[the so-called post-Born corrections:][]{ch02,hs03}.
In summary, we transferred the complex polarization field $(Q+iU) (\bftheta)$ to the new field $(\tilde{Q}+i\tilde{U}) (\bftheta)$ using the deflection angle $\bfd$ and the rotation of polarization $\beta$ as \citep[e.g.,][]{maro16,lhc17}
\beq
\left( \tilde{Q}+i \tilde{U} \right)(\bftheta) = e^{-2 i \beta(\bftheta)} \left( Q+iU \right)(\bftheta+\bfd),
\label{QU_trans}
\eeq
where $Q$ and $U$ are the Stokes parameters.
We constructed $108$ lensed CMB maps for each resolution $N_{\rm side}=4096$ and $8192$.

Analytic expressions of polarization rotation $\beta$ were derived by \cite{maro16} and \cite{lhc17} based on the perturbative expansion of gravitational potential but with different approximation schemes.  
\cite{maro16} used a post-Born approximation with a leading correction from multiple-lensing deflections and found that $\beta$ is the same as the field rotation $\omega$ given in Equation (\ref{jacob_mat}), resulting in $\beta$ of order $10^{-3}$ rad.
\cite{lhc17}, on the other hand, used the Born approximation, but took into account corrections from multiple-lensing deflections and the emission-angle effect, and found that the rotation angle is quadratic in the deflection angle and is very small ($\beta \approx 10^{-6}$ rad). 
In what follows, we consider two cases: one is $\beta=0$ as a default, and the other is $\beta=\omega$.\footnote{We set $\beta=\omega$ in the previous version of our draft (arXiv:1706.01472v1). The field rotation $\omega$ at the last scattering surface can be numerically obtained from the ray-tracing simulation.}
This difference will affect only the CMB B-mode power spectrum by a few percent for the very small scale $\ell > 2000$ (Section 3.5).

The bottom panel of Figure \ref{fig_sample_maps} shows a contour plot of the amplitude of the CMB lensing B-mode.
The polarization patterns can be decomposed into E-mode (positive parity) and B-mode (negative parity).
The \texttt{HEALPix} scheme (\texttt{map2alm}) automatically evaluates the multipole coefficients $a_{\ell m}$ of the polarization for the E- and B-modes separately from the full-sky map.
We can then transform the B-mode coefficient to the contour map using the multipole expansion. 
A typical B-mode patterns size in the panel, as determined by the peak of the B-mode power spectrum, is a few $10$ arcmin (we calculate the angular power spectra of the temperature and polarization fluctuations in Section 3.4).

\section{Basic Statistics}

In this section, we present detailed comparisons between our simulation results and theoretical predictions for the convergence power spectrum (Section 3.1), shear correlation functions (Section 3.2), halo-galaxy lensing (Section 3.3), angular correlation function of halos (Section 3.4), and finally CMB power spectra (Section 3.5).
We will show the measurements with two resolutions, $N_{\rm side}=4096$ and $8192$, to clearly show the angular scales that are resolved in each $N_{\rm side}$.

\subsection{Convergence Power Spectrum}

\begin{figure}
\hspace*{-0.5cm}
\includegraphics[width=1.2\columnwidth]{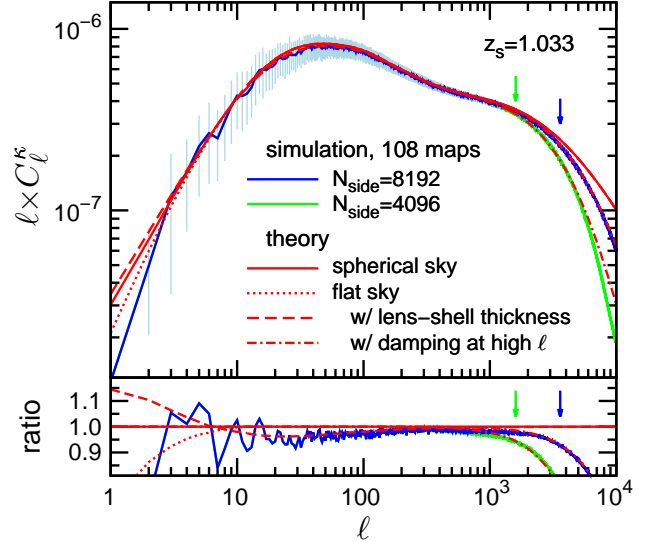}  
\caption{Convergence power spectrum $\ell \times C_\ell^\kappa$ as a function of the multipole $\ell$ at source redshift $z_{\rm s}=1.033$. 
  The blue and green curves show the average power spectra of the $108$ maps for angular resolutions $N_{\rm side} =8192$ and $4096$, respectively.
  The light-blue bars show the standard deviations of the $108$ maps. 
  The solid (dotted) red curve is the theoretical prediction of the spherical-sky (flat-sky) formula.
  The dashed curve takes into account the effects of the finite thickness of the lens shells (see Appendix B for a detailed discussion).
  The dot-dashed curves include a damping factor at high $\ell$ arising from the finite angular resolution.
  The down arrows indicate the scale at which the simulation results converge within $5\%$.
  The bottom panel shows ratios to the solid red curve.
}
\vspace*{0.3cm}
\label{fig_cl_conv}
\end{figure}

Images of distant sources are distorted by the foreground mass distribution.
This distortion is usually characterized by convergence and shear components. 
The convergence field at a source redshift $z_{\rm s}$ is, at the lowest order, given by the weighted surface mass density as \citep[e.g.,][]{bs01}
\beq
\kappa(\bftheta) = \frac{3 H_0^2 \Omega_{\rm m}}{2} \int_0^{z_{\rm s}} \!\! 
\frac{dz}{H(z)} \frac{r(z) \left( r_{\rm s}-r(z) \right)}{a(z) \, r_{\rm s}}
\, \delta (r \bftheta,r;z),
\label{eq_conv}
\eeq
where $H(z)$ is the Hubble expansion rate at $z$, $a(z)=1/(1+z)$ is the scale factor, $r(z)$ is the comoving distance to the redshift $z$, $r_{\rm s} (=r(z_{\rm s}))$ is the distance to the source, and $\delta(\bfx;z)$ is the density contrast at position $\bfx$ at $z$. Here, we decompose the position $\bfx$ into the radial distance $r$ and the perpendicular distance $r \bftheta$ in Equation (\ref{eq_conv}).
The convergence power spectrum in the flat-sky approximation is given by
\beq
C_\ell^\kappa = \frac{9 H_0^4 \Omega_{\rm m}^2}{4} 
\int_0^{z_{\rm s}} \!\! \frac{dz}{H(z)} \frac{\left( r_{\rm s} -r(z) \right)^2}{a(z)^2 \, r_{\rm s}^2} P_\delta \! \left( k=\frac{\ell}{r(z)};z \right),
\label{cl_flatsky}
\eeq
where $\ell$ is the multipole and $P_\delta(k;z)$ is the power spectrum of the density contrast at wavenumber $k$ at redshift $z$.
In this expression, we use the revised Halofit formula for the nonlinear matter power spectrum \citep{s03,t12}.

Figure \ref{fig_cl_conv} shows the average convergence power spectrum at $z_{\rm s}=1.033$ evaluated from the $108$ maps.
The blue and green curves correspond to the resolutions at $N_{\rm side}=8192$ and $4096$, respectively (in this plot, we do not bin the results).
We used the \texttt{map2alm} and \texttt{alm2cl} \texttt{HEALPix} schemes to measure the power spectra from the maps.
The dotted red curve shows the theoretical prediction in the flat-sky approximation given in Equation (\ref{cl_flatsky}).
We also consider a correction of the sky curvature at large angular scales, with the solid red curve including the correction proposed in \citet{hu00} assuming a linear power spectrum of gravitational potential.
The theoretical prediction of cosmic shear power spectra on the sphere includes the density power spectrum between different cosmological epochs for a given wavenumber, indicating that some approximations are required to evaluate such unequal-time cross spectra \citep[e.g., see][for details\footnote{ \cite{kilbinger17} showed that the flat-sky power spectrum (\ref{cl_flatsky}) was valid within $5\% (1\%)$ for $\ell>5(20)$ for the CFHTLenS source redshift distribution (see their Figure 1).}]{kitching16, kilbinger17}.
As the spherical-sky formula in \citet{hu00} is valid only in the linear regime, we simply connect the spherical-sky formula to the flat-sky formula at an intersection point\footnote{The spherical-sky (flat-sky) formula predicts larger power at larger (smaller) scales. They intersect at $\ell=13$ in our setting.} in plotting the solid curve.
Although the spherical-sky formula gives $38\%$, $16\%$, and $9\%$ higher amplitudes than the flat-sky formula for $\ell=1,2$, and $3$, respectively, the difference is smaller than $5\% \, (1\%)$ for $\ell \geq 4 \, (8)$.
The differences between the two formulae are smaller than the standard deviations of the $108$ samples at low multipoles (shown as the light-blue bars).
The bottom panel shows the ratio of the simulation results to the theoretical prediction of the spherical-sky formula.
For very small scales ($\ell > 10^3$), the simulation results underestimate the power owing to a lack of angular resolution.
The dot-dashed curves take the finite angular resolution into account based on a simple damping factor at small scales 
\beq
C_\ell^\kappa \rightarrow \frac{C_\ell^\kappa}{1+\left( \ell/\ell_{\rm res} \right)^2}.
\label{damping}
\eeq
By setting a damping scale $\ell_{\rm res}=1.6 \times N_{\rm side}$, the dot-dashed curve closely fits the simulation results.
At intermediate scales ($\ell \simeq 10-100$), the simulation results are slightly lower than  those of the theoretical model shown in the bottom panel (but the difference is less than $5 \%$).
This is because the finite thickness of the lens shells influences the angular power spectrum of surface density fluctuations on a shell (a detailed discussion of this is given in Appendix B).
To take this effect into account, we convolved the matter power spectrum with the window function of the shell.
The dashed curve shows the results of replacing the power spectrum $P_\delta(k,z)$ in Equation (\ref{cl_flatsky}) by the convolved spectrum $P^W_\delta(k,z)$ in Equation (\ref{pk_replace2}). 
This replaced curve agrees closely with the simulation results.
In summary, our simulation results agree with theoretical prediction within $5\%$ at $\ell < 3000$ $(1400)$ for $N_{\rm side}=8192$ $(4096)$.

We further checked that the convergence power spectrum at $N_{\rm side}=8192$ $(4096)$ agrees with the higher-resolution result at $N_{\rm side}=16384$ within $5\%$ for $\ell<3600$ $(1600)$ at $z_{\rm s}=0.3-5$.
The down blue and green arrows in Figure \ref{fig_cl_conv} indicate the scales at which the simulation results converge.

\subsection{Shear Correlation Functions}

\begin{figure}
\includegraphics[width=1.1\columnwidth]{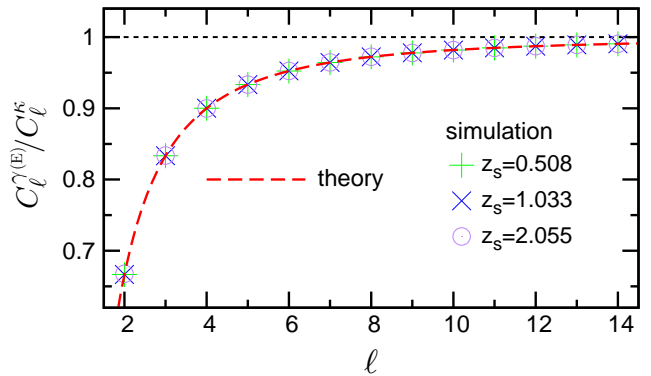} 
\vspace*{-1.cm}
\caption{
  Ratio of $C_\ell^{\gamma ({\rm E})}$ to $C_\ell^\kappa$ for $z_{\rm s}=0.508$ (green pluses), $1.033$ (blue crosses), and $2.055$ (purple circles).
  Here, $C_\ell^{\gamma ({\rm E})}$ and $C_\ell^\kappa$ are the averages measured from the $108$ maps with $N_{\rm side}=8192$.
  The dashed red curve denotes the theoretical prediction in the spherical sky, given in Equation (\ref{cl_kappa_E}).
}
\label{fig_cl_conv-EE_ratio}
\end{figure}

\begin{figure*}
\includegraphics[width=2.\columnwidth]{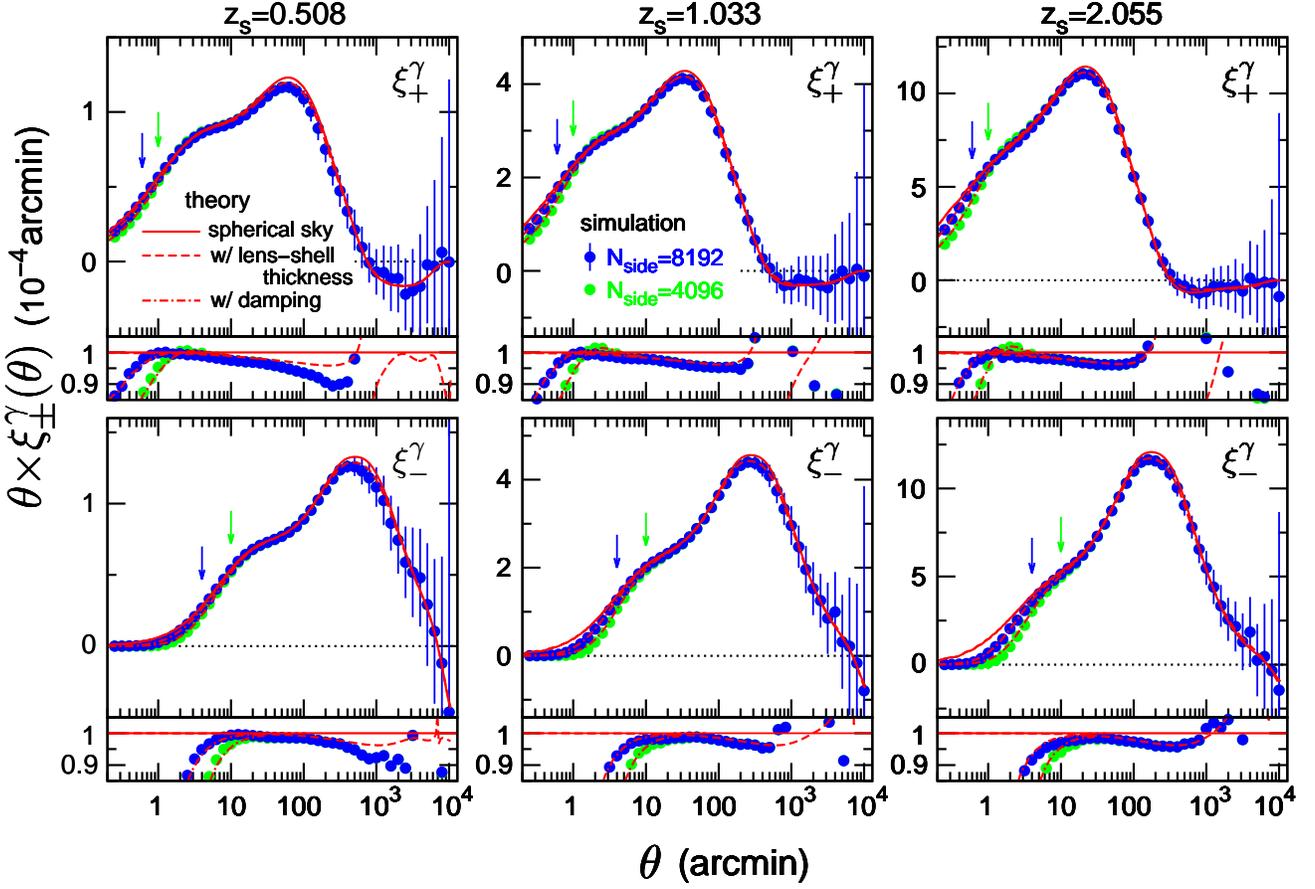} 
\caption{Shear correlation functions $\xi^\gamma_\pm (\theta)$ as a function of $\theta$ at source redshifts $z_{\rm s}=0.508$, $1.033$, and $2.055$ (left to right panels, respectively).
  The upper (lower) panels plot $\xi_+^\gamma$ $(\xi_-^\gamma)$.
  The $y$-axes give $\theta \times \xi_{\pm}^\gamma (\theta)$ in units of $10^{-4}$ arcmin.
  The blue and green filled circles are the averages of the $108$ maps with
  $N_{\rm side}=8192$ and $4096$, respectively.
  The blue error bars show the standard deviations.
  The solid red curves show the theoretical prediction in the spherical-sky formula.
  The dashed and dot-dashed curves include corrections for lens-shell thickness and finite angular resolution, respectively.
  The down blue and green arrows indicate the scales at which the simulation results converge within $5\%$.
  The small bottom panels show ratios to the solid red curves.
}
\label{fig_xi_pm}
\end{figure*}

We next discuss the correspondence between our results and theoretical values for the shear correlation functions.
The shear components $\gamma_{1,2}$ are defined in the \texttt{HEALPix} spherical coordinates in the standard manner, where $\gamma_1$ is a component along a line of longitude (from the north pole to the south pole) and $\gamma_2$ is a component rotated by $45^\circ$ (see also Figure 5 of the \texttt{HEALPix} primer\footnote{http://healpix.sourceforge.net/documentation.php} for the coordinate convention).
The shear can be decomposed into tangential and cross modes in terms of the relative angular position between two points.
The tangential mode $\gamma_{\rm t}$ is defined along a great arc connecting two points, while the cross mode $\gamma_{\rm \times}$ is a mode rotated by $45^\circ$.
The correlation of the shears at two angular positions $\bftheta_{\rm i}$ and $\bftheta_{\rm j}$ is then given by \citep[e.g.,][]{k92,s02}
\beq
\xi_\pm^\gamma(\theta) = \langle \left[ \gamma_{\rm t}(\bftheta_{\rm i})
  \gamma_{\rm t}(\bftheta_{\rm j}) \pm \gamma_{\rm \times}(\bftheta_{\rm i})
  \gamma_{\rm \times}(\bftheta_{\rm j}) \right] \rangle,
\eeq
where $\theta$ is the angular separation between $\bftheta_{\rm i}$ and $\bftheta_{\rm j}$ \citep[see also][Section 3.2, for a discussion about angular correlation functions in spherical coordinates]{k13}.
Note that the $\xi_{\pm}^\gamma(\theta)$ are independent of the coordinate system and $\xi_+^\gamma(\theta)$ is the same as the convergence correlation function in the flat-sky approximation.

The relation between the shear correlation functions and the angular power spectra in spherical coordinates is discussed in detail in previous works \citep[e.g.,][]{s96,hw97,kks97,zs97,nl99,hu00}.
The shear power spectrum is generally decomposed into E- and B-modes. 
According to the previous works, the shear correlation functions $\xi^\gamma_\pm (\theta)$ can be rewritten in terms of the multipole expansion of the E/B-mode shear power spectra $C_\ell^{\gamma ({\rm E/B})}$ as
\beq
\xi_\pm^\gamma (\theta) = \sum_{\ell=2}^\infty \sqrt{\frac{2 \ell+1}{4 \pi}}
\left[ C_\ell^{\gamma({\rm E})} \pm C_\ell^{\gamma({\rm B})} \right]
{}_2Y_{\ell \mp 2}(\theta,0),
\label{xi_pm_multiexp}
\eeq
where ${}_2Y_{\ell m}(\theta,\phi)$ are the spin-2 spherical harmonics.
The shear E-mode and convergence power spectra are related via \citep[e.g.,][]{b13}
\beq
C_\ell^{\gamma ({\rm E})} = \frac{(\ell+2) (\ell-1)}{\ell (\ell+1)} C^\kappa_\ell.
\label{cl_kappa_E}
\eeq
Figure \ref{fig_cl_conv-EE_ratio} plots the ratio $C_\ell^{\gamma ({\rm E})}/C^\kappa_\ell$ measured from the full-sky maps (denoted by the colored symbols) compared to the theoretical prediction of $[(\ell+2) (\ell-1)]/ [\ell (\ell+1)]$ (the dashed red curve), showing that this relation (\ref{cl_kappa_E}) holds in our simulation.
Note that in the small-angle limit, ${}_2Y_{\ell m}(\theta,\phi)$ is approximated by ${}_2Y_{\ell \mp 2}(\theta,0) \simeq \sqrt{\ell/(2 \pi)} J_{0/4}(\ell \theta)$, where $J_{0/4}(x)$ is the zeroth-order or fourth-order Bessel function.
In this case, Equation (\ref{xi_pm_multiexp}) reduces to the flat-sky formula
\beq
\xi_\pm^\gamma (\theta) \simeq \frac{1}{2 \pi} \int_0^\infty \!\! d \ell \, \ell \left[ C_\ell^{\gamma({\rm E})} \pm C_\ell^{\gamma({\rm B})} \right] J_{0/4}(\ell \theta).
\label{xi_pm_flatsky}
\eeq

Figure \ref{fig_xi_pm} shows, from left to right, the shear correlation functions at $z_{\rm s} = 0.508$, $1.033$, and $2.055$.
The top (bottom) panels show $\xi_+^\gamma$ $(\xi_-^\gamma)$.
The filled circles with error bars are the averages with standard deviations of the $108$ maps.
The data points are binned with $\Delta \log [\theta({\rm arcmin})] = 0.1$.
To measure the correlation functions from the maps, we employed the multipole expansion of the power spectra in Equation (\ref{xi_pm_multiexp}).
We evaluated the shear E- and B-mode power spectra ($C_\ell^{\gamma({\rm E})}$ and $C_\ell^{\gamma({\rm B})}$) from the maps using the \texttt{HEALPix} scheme and then computed the summation (\ref{xi_pm_multiexp}) up to $\ell_{\rm max}=3 N_{\rm side}$.
Here we do not include the shape noise for the analysis of the shear maps.
The solid red curves show the theoretical prediction given in Equations (\ref{xi_pm_multiexp}) and (\ref{cl_kappa_E}).
Here, we adopt the spherical-sky formula for $C^\kappa_\ell$ (see Section 3.1), but simply assume $C_\ell^{\gamma ({\rm B})}=0$ because the B-mode $C_\ell^{\gamma ({\rm B})}$ is approximately three orders of magnitude smaller than the E-mode $C_\ell^{\gamma ({\rm E})}$ at $\ell<10^4$ \citep{ch02,kh10}.
The dashed and dot-dashed curves include the effects of lens-shell thickness and finite angular resolution, respectively.
These curves are the same as in Figure \ref{fig_cl_conv}, but simply transformed into the correlation functions using the multipole expansion (\ref{xi_pm_multiexp}).
The small bottom panels show ratios to the spherical-sky theoretical model.
As seen in the small bottom panels, the simulation results are slightly smaller ($\lesssim 5\%$) than the theoretical model at intermediate scales ($\theta \sim 10-1000$ arcmin), although the difference can be attributed to the shell-thickness effect (the dashed curves).
The simulation results at $z_{\rm s}=0.508$ show slightly more suppression than the theoretical prediction (the dashed curves) at intermediate scales ($\theta \sim 100-1000$ arcmin) because the theoretical model including the shell-thickness effect is less accurate in nearby lens planes at $z < 0.1$ (see also Appendix B).
The scatter in the simulation results at large scales ($\theta >$ several $100$ arcmin) is caused by sample variance.
The suppressions of $\xi_\pm^\gamma$ at small scales ($\theta \lesssim 1-10$ arcmin) arise from the lack of angular resolution, which can be fitted well by including the damping factor (\ref{damping}) (the dot-dashed curves). 
In particular, the suppressions are significant for $\xi_-^\gamma$ compared to $\xi_+^\gamma$, because $\xi_-^\gamma$ is more sensitive to higher multipoles of the power spectrum (the spherical harmonics $_2Y_{\ell 2}(\theta,0)$ have a first peak at higher $\ell$ than $_2Y_{\ell -2}(\theta,0)$ for a given $\theta$) in Equation (\ref{xi_pm_multiexp}), and therefore $\xi_-^\gamma$ is more influenced by this suppression at large $\ell$.
In summary, the simulation results for $\xi_+^\gamma$ and $\xi_-^\gamma$ at $N_{\rm side}=8192$ $(4096)$ agree with the theoretical prediction within $5\%$ at $\theta > 0.6$ $(1)$ arcmin and $\theta > 6$ $(10)$ arcmin, respectively.

We also confirmed that the shear correlation functions $\xi_+^\gamma$ and $\xi_-^\gamma$ at $N_{\rm side}=8192$ $(4096)$ agree with those at the higher resolution $N_{\rm side}=16384$ within $5\%$ for $\theta>0.6 \, (1.0)$ arcmin and for $\theta>4 \, (10)$ arcmin, respectively, within the redshift range $z_{\rm s}=0.3-5$.
The down blue and green arrows in Figure \ref{fig_xi_pm} indicate the respective scales at which the convergence occurs.

\subsection{Halo-galaxy Lensing}

\begin{deluxetable}{cccc}
\tablecaption{Halo Samples}
\startdata
  \hline 
 Name & Mass ($h^{-1} {\rm M}_\odot$) & Redshift & Number of Halos \\
\hline
 sample 1  & $M_{200 {\rm b}}>10^{13}$  & $0.47<z<0.59$ & $2.2 \times 10^6$ \\
 sample 2  & $M_{200 {\rm b}}>10^{13.5}$  & $0.16<z<0.36$ & $3.6 \times 10^5$ \\
 sample 3  & $M_{200 {\rm b}}>10^{14}$  & $0.10<z<0.33$ & $5.8 \times 10^4$ 
\enddata
\tablecomments{
  The halo sample name, the minimum halo mass, the redshift range, and the number of halos in the all-sky sample. Sample 1,2, and 3 roughly correspond to the CMASS galaxy-like, LRG-like, and redMaPPer cluster-like samples.}
\label{halo_samples}
\end{deluxetable}

\begin{figure*}
\includegraphics[width=2.\columnwidth]{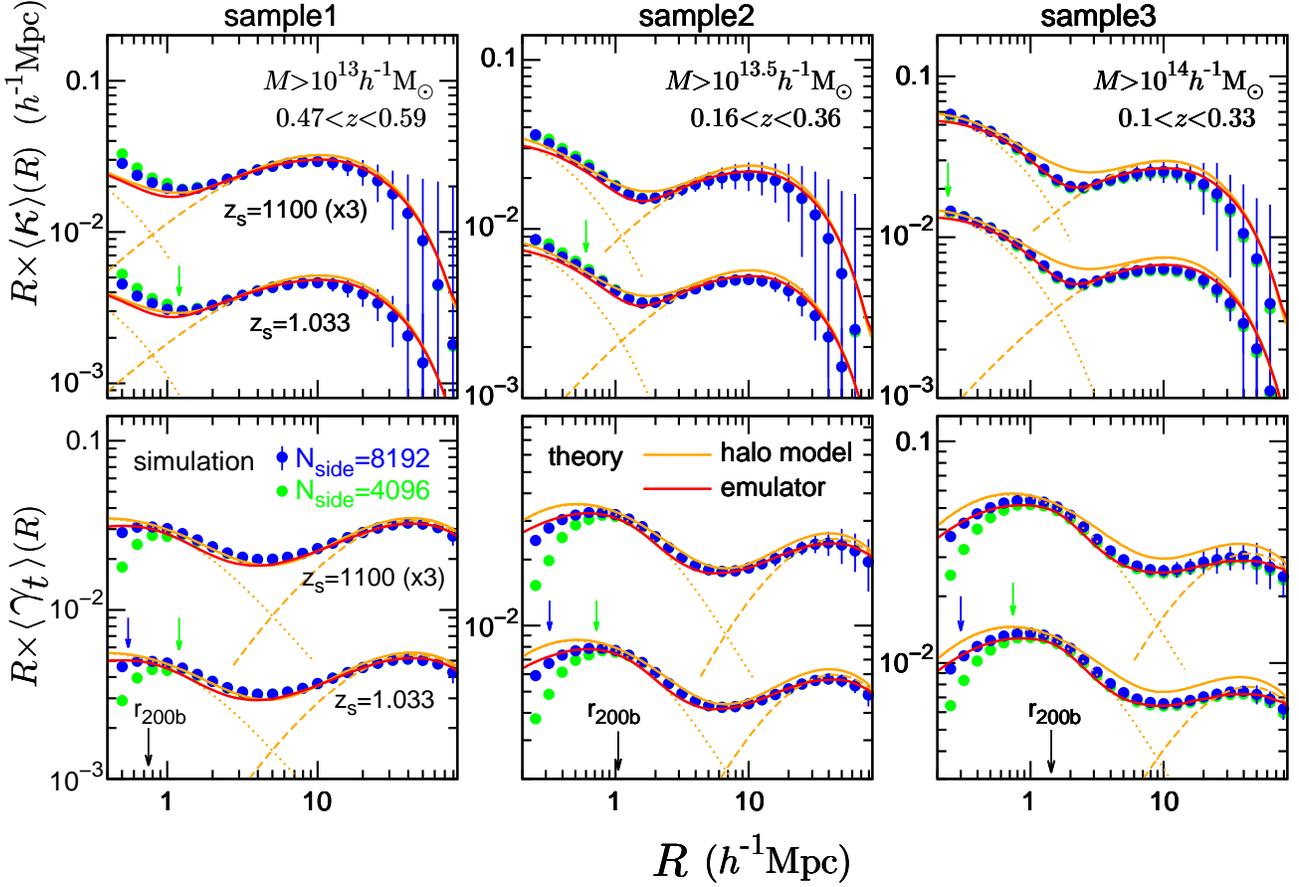} 
\caption{
  Mean convergence (upper panels) and tangential shear (lower panels) profiles of halos for the three halo samples described in Table \ref{halo_samples}.
  The $x$-axes are the (comoving) projected separations $R$ from the halo centers, while the $y$-axes are the mean profiles. 
  The source redshifts are $z_{\rm s}=1.033$ and $1100$, where the results with $z_{\rm s}=1100$ are multiplied by a factor three for clearer presentation.
  The blue and green symbols are the average simulation results of the $108$ maps with a bin width of $\Delta \log [R/(h^{-1} {\rm Mpc})]=0.1$ at $N_{\rm side}=8192$ and $4096$, respectively, with the blue bars showing the standard deviations.
  The dotted (dashed) orange curves are the one-halo (two-halo) terms in the halo model, while the solid orange curves show their sums.
  The solid red curves indicate the theoretical prediction of \texttt{the dark emulator} (T. Nishimichi et al. 2017, in preparation).
  The down black arrows in the lower panels indicate the mean halo radius, $r_{200 {\rm b}}$, for each sample. 
  The down blue and green arrows indicate the scales at which the simulation results converge to within $5\%$ accuracy.
}
\label{fig_gg_lens}
\end{figure*}

In this subsection, we calculate the mean convergence and tangential shear profiles of halos.
The cross correlation of foreground galaxies with background shear can provide the excess surface mass density of foreground galaxies (galaxy-galaxy lensing).
Combining galaxy-galaxy lensing with the galaxy clustering enables the inference of galaxy bias, which helps to break the degeneracy between cosmological parameters \citep[e.g.,][]{sel05}.
In addition, it is also possible to measure the mean density profile of clusters in this manner.

The \cite{planck16xv} recently provided a convergence map reconstructed from temperature and polarization data.
The SPT collaboration also published a full-sky convergence map reconstructed from SPT-SZ and Planck temperature maps \citep{omori17}.
Several authors subsequently measured the mean mass profile of galaxies by correlating the foreground galaxies with the background convergence field \citep[e.g.,][]{m17,smb17}.
The Planck convergence map has also been used to study the cross-correlation measurement with galaxy density \citep[e.g.,][]{gian16}.

Here, we calculate for demonstration purposes the mean convergence and tangential shear profiles of three halo samples in Table \ref{halo_samples} for two sources of background galaxies at $z_{\rm s}=1.033$ and the CMB at $z_{\rm s}=1100$.
The table summarizes the minimum halo mass, redshift range, and number of the halos for each sample.
Samples 1,2, and 3 roughly correspond to the SDSS CMASS galaxy-like sample \citep{a15,m15}, the LRG-like sample \citep{r10,m13}, and the redMaPPer cluster-like sample \citep{r14,m16}, respectively.
We did not include subhalos in this analysis and also ignored the shape noise of the background field.
Using the halo catalogs (see Section 2.3), we directly calculated the mean profiles of the foreground halos.
The mean convergence profile at a separation $R$ from the halo center can be written as
\beq
\langle \kappa \rangle (R) = \frac{1}{N_{\rm h}} \sum_{{\rm i}=1}^{N_{\rm h}} \, \frac{1}{N_{\rm pix(i)}} \!\!\! \sum_{{\rm j}=1}^{N_{\rm pix(i)}} \kappa(R_{\rm ij}),
\eeq
where $R$ is the projected comoving separation, $N_{\rm h}$ is the number of halos in the sample, $R_{\rm ij}$ is the separation of the $i$th halo from the $j$th background convergence field, and $N_{\rm pix(i)}$ is the number of pixels in the convergence map around the $i$th halo in an annulus with radius $R-\Delta R/2 < R_{\rm ij} < R+\Delta R/2$. 
The mean tangential shear profile $\langle \gamma_{\rm t} \rangle (R)$ is the same, but with $\kappa$ simply replaced by $\gamma_{\rm t}$ in the above equation.
We calculated the profiles using a brute-force direct summation of all pairs in the $108$ maps.

To check our simulation results, we adopted two theoretical models of the halo model \citep[e.g.,][]{cs02} and the \texttt{dark emulator} (T. Nishimichi et al. in preparation).
We briefly explain the halo model considered here for comparison. 
Essentially, we followed the halo model parameters employed in \cite{oh11} \citep[see also][]{ot11}, which can reproduce the simulation results of the mean convergence profile very well \citep{sato09}.
The mean convergence at a (comoving) separation of $R$ can be formally written as
\beq
\langle \kappa \rangle (R) = \frac{1}{N_{\rm h}} \int_{z_1}^{z_2} \!\! dz \frac{dV}{dz}
\int_{M_{\rm min}}^{\infty}
\!\!\!\! dM \frac{dn}{dM}(M,z) \, \kappa(R;M,z),
\label{mean_kappa_hg}
\eeq
with the number of halos
\beq
N_{\rm h} = \int_{z_1}^{z_2} \!\! dz \frac{dV}{dz} \int_{M_{\rm min}}^{\infty} \!\!\!\! dM \frac{dn}{dM}(M,z),
\nonumber
\eeq
where $dV/dz$ $(= 4 \pi r(z)^2 / H(z))$ is the comoving volume element, $z_1$ and $z_2$ are the redshift range of the halos, $M_{\rm min}$ is the minimum halo mass given in Table \ref{halo_samples}, and $dn/dM$ is the halo mass function for which we employ the fitting formula in \cite{t08}.
Here, we use the mass contained within the spherical overdensity region with a density $200$ times higher than the mean (comoving) background density $\bar{\rho}_{\rm m}$, i.e., $M_{200 {\rm b}}=200 \bar{\rho}_{\rm m} \times (4 \pi r_{200 {\rm b}}^3)/3$, where $r_{200 {\rm b}}$ is the (comoving) halo radius.
The convergence $\kappa(R;M,z)$ in Equation (\ref{mean_kappa_hg}) comprises two terms: the halo density profile (the so-called ``one-halo term''), and the matter distribution around the halo (the so-called ``two-halo term'').
For the density profile, we adopted an NFW profile with a truncation at large radius \citep{nfw97,bmo09}
\beq
\rho(r) = \frac{\rho_{\rm s}}{(r/r_{\rm s})(1+r/r_{\rm s})^2} \frac{1}{\left[ 1+(r/r_{\rm t})^2 \right]^2},
\label{tnfw}
\eeq
where the truncation radius was set to $r_{\rm t}=3 \, r_{200{\rm b}}$.
To determine the characteristic scale $r_{\rm s}$, we employed a fitting formula of the concentration parameter $c=r_{200 {\rm b}}/r_{\rm s}$ calibrated by an $N$-body simulation \citep{d08}.
The characteristic density $\rho_{\rm s}$ is determined by using Equation (\ref{tnfw}) to determine the mass inside $r_{200 {\rm b}}$ as $M_{200 {\rm b}}=4 \pi \int_0^{r_{200 {\rm b}}} dr \, r^2 \rho(r)$.
Note that $\rho_s$ and $r_{\rm s}$ depend on both $M$ and $z$.
Based on the halo surface mass density of $\Sigma(R)=\int_{-\infty}^{\infty} dl \rho[(R^2+l^2)^{1/2}]$, the one-halo term is given by the surface density divided by the critical density, $\kappa_{\rm 1h}(R;M,z)=\Sigma(R;M,z)/\Sigma_{\rm crit}(z)$ with $\Sigma_{\rm crit}(z)=(1+z) r_{\rm s}/[ 4 \pi r(z) (r_{\rm s}-r(z)) ]$.

The two-halo term represents the cross correlation between the halo and the surrounding matter,
\beq
\kappa_{\rm 2h}(R;M,z) = \frac{b_{\rm h}(M,z) \bar{\rho}_{\rm m}}{\Sigma_{\rm crit}(z)} \int \frac{k dk}{2 \pi} J_0(k R) P_{\rm m,lin}(k;z),
\eeq
where $P_{\rm m,lin}$ is the linear matter power spectrum (calculated by CAMB) and $J_0(x)$ is the zeroth-order Bessel function. 
We used the fitting formula of the halo bias $b_{\rm h}(M,z)$ in \cite{t10}.
The convergence was calculated as the sum of the one- and two-halo terms: $\kappa(R;M,z)=\kappa_{\rm 1h}(R;M,z) + \kappa_{\rm 2h}(R;M,z)$.
Inserting these terms into Equation (\ref{mean_kappa_hg}) produced the mean convergence profile of the halos. 

The tangential shear profile is similar to the convergence profile, but simply given by \citep[e.g.,][Section 2.3 of Part 3]{skw06}
\beq
\langle \gamma_{\rm t} \rangle(R) = \langle \bar{\kappa} \rangle (<R)- \langle \kappa \rangle(R),
\eeq
where $\langle \bar{\kappa} \rangle(<R)$ is the mean convergence profile inside $R$, $\langle \bar{\kappa} \rangle (<R)=(2/R^2) \int_0^R dR^\prime R^\prime \, \langle \kappa \rangle (R^\prime)$.

We also employed \texttt{the dark emulator} constructed on a series of $N$-body simulations, which will soon be publicly available (T. Nishimichi et al., in preparation).
Cosmological $N$-body simulations with $2048^3$ particles were performed for $100$ six-parameter $w$CDM cosmological models sampled around the Planck 2015 best-fit flat $\Lambda$CDM cosmology \citep{planck16xiii} with 21 outputs dumped over a redshift range of $z=0-1.5$.
An efficient sampling scheme based on the maxi-min distance Latin Hypercube design and the use of functional interpolation based on Gaussian processes allowed for an accurate prediction of the statistical quantities with a relatively small number of sampled points (see e.g., \citealt{emu1,emu2,emu3} for earlier studies on the nonlinear matter power spectrum and \citealt{emu4} for recent developments by the Mira-Titan Universe project).
\texttt{The dark emulator} focuses on the halo clustering properties and provides the halo mass function, and auto- and cross-correlation functions of the halo and matter fields.
The correlation functions are measured using a hybrid direct-FFT algorithm to achieve both accuracy (on scales smaller than the FFT grid) and speed.
The excess surface mass density of halos can be obtained from the halo-mass cross-correlation function by convolving the appropriate kernel with a quick implementation of the Fourier transform (FFTLog, \citealt{hamilton00}).
T. Nishimichi et al. (in preparation) developed a python package in which the hyper-parameters for the Gaussian processes and the data table are prepared to enable a quick evaluation of the related statistical quantities for a given redshift, halo mass, and cosmological parameters.
The current version presented here is based on 40 simulations with a box size of $1\,h^{-1}\mathrm{Gpc}$ at 40 different cosmological parameters.
The statistics are derived for central halos identified by \textsc{Rockstar}.

Figure \ref{fig_gg_lens} shows, from left to right, the mean convergence (upper panels) and tangential shear profile (lower panels) for halo samples 1 to 3. 
The source redshifts are $z_{\rm s}=1.033$ and $1100$.
The filled blue and green circles are the averaged results of the $108$ maps at $N_{\rm side}=8192$ and $4096$, respectively.
The error bars show the standard deviations of the $108$ maps.
We excluded the subhalos from the samples.
The solid orange and red curves show the theoretical predictions produced by the halo model and \texttt{the dark emulator}, respectively. 
The dotted and dashed orange curves show the one- and two-halo terms in the halo model.
\texttt{The dark emulator} provides closer agreement, particularly at the intermediate scale between the one- and two-halo regimes.
For smaller scales of $R \lesssim 1h^{-1}$Mpc, the simulation results deviate from the theoretical models owing to the finite angular resolution.
This corresponds to angular scales of $\sim \, 3 \, {\rm arcmin} [R/(h^{-1} {\rm Mpc})] [r/(h^{-1} {\rm Gpc})]^{-1}$ (where $r$ is the distance to the halo), which are comparable to the angular resolution of the shear correlation functions shown in Figure \ref{fig_xi_pm}.
The convergence profiles show a damping feature at larger scales ($R \gtrsim 30 \, h^{-1}$Mpc) compared to the theoretical predictions.
This is partially a result of the lens-shell thickness effect underestimating the correlation.
The separation $R \approx 30 \, h^{-1}$Mpc corresponds to an angular scale $\theta \approx 100$ arcmin, at which the suppression seen in Figure \ref{fig_xi_pm} is significant.
In summary, the simulation results at $N_{\rm side}=8192$ agree with both theoretical models within $30 \%$ for the convergence profiles at $0.5 \leq R/(h^{-1}{\rm Mpc}) \leq 40$ and for the tangential shear profiles at $0.4 \leq R/(h^{-1}{\rm Mpc}) \leq 100$.

The down blue and green arrows in Figure \ref{fig_xi_pm} indicate the innermost scales at which the mean profiles for $\langle \kappa \rangle$ and $\langle \gamma_{\rm t} \rangle$ at $N_{\rm side}=8192$ $(4096)$ agree with the corresponding higher resolution $N_{\rm side}=16384$ profiles within $5\%$.

\subsection{Halo Clustering}

\begin{figure}
\hspace*{-0.5cm}
\includegraphics[width=1.2\columnwidth]{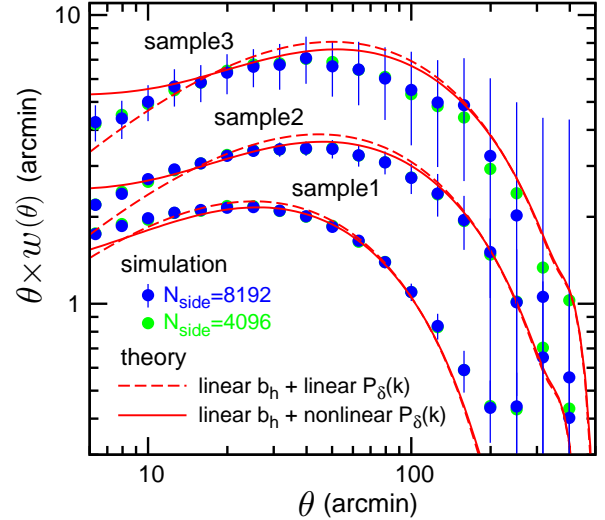} 
\caption{Halo angular correlation functions for samples from 1 to 3 (in Table \ref{halo_samples}), shown bottom to top, respectively. The blue and green circles with error bars are the averages with standard deviations of the $108$ maps at $N_{\rm side}=8192$ and $4096$, respectively.
The solid (dashed) red curves show the theoretical prediction in Equation (\ref{w_halo}) assuming linear halo bias and a nonlinear (linear) matter power spectrum.
}
\label{fig_w_halo}
\end{figure}

In this subsection, we calculate the angular two-point correlation function of the halo samples in Table \ref{halo_samples}.
The number density contrast of halos $\delta_{\rm h}(\bfx;M,z)$ and the matter density contrast $\delta(\bfx;z)$ are related via $\delta_{\rm h}(\bfx;M,z)=b_{\rm h}(M,z) \delta(\bfx;z)$.
Then, the projected halo number density contrast is given by
\beqa
 \delta_{\rm h}^{\rm 2D}(\bftheta) = \frac{1}{N_{\rm h}} \int_{z_1}^{z_2} \!\! dz \frac{dV}{dz} \int_{M_{\rm min}}^\infty \!\!\! dM \frac{dn}{dM}(M,z)  \nonumber \\
 \times \, b_{\rm h}(M,z) \, \delta(r \bftheta,r;z),
 \eeqa
 where the term $dn/dM$ is necessary to take into account the halo mass distribution in each sample \citep[e.g.,][]{cs02}. 
The angular two-point halo correlation can be obtained from the above equation under the Limber approximation as \citep[e.g.,][]{dod03}
\beqa
w(\theta) &=& \langle \delta_{\rm h}^{\rm 2D} (\bftheta+\bftheta^\prime) \delta_{\rm h}^{\rm 2D} (\bftheta^\prime) \rangle
\nonumber \\
 &=& \frac{8 \pi}{N_{\rm h}^2} \int_{z_1}^{z_2} \!\! \frac{dz}{H(z)} \left[ \, r(z) \! \int_{M_{\rm min}}^\infty \!\!\! dM \frac{dn}{dM}(M,z) \, b_{\rm h}(M,z) \right]^2   \nonumber \\
 && ~~~~~ \times \int \! d\ell \, \ell \, P_\delta \left( k=\frac{\ell}{r(z)};z \right) J_0(\ell \theta).
\label{w_halo}
\eeqa
where $P_\delta(k;z)$ is the matter power spectrum.
We then calculated the halo correlation function $w(\theta)$ for the three halo samples in Table \ref{halo_samples}.

Figure \ref{fig_w_halo} plots the angular correlation functions of halos from the three samples.
The blue and green dots with error bars are the averages with standard deviations of the $108$ maps at $N_{\rm side}=8192$ and $4096$, respectively.
In this case, we first calculated the halo angular power spectra from the maps, subtracted the shot noise, and finally transformed the results into angular correlation functions using the multipole expansion.
The solid (dashed) red curves show the theoretical prediction (Equation \ref{w_halo}) with the nonlinear (linear) matter power spectrum.
We employed the revised Halofit to derive the nonlinear matter power spectrum and the Tinker fitting functions to derive the mass function and linear halo bias, as was done in Section 3.3.
The error bars for sample 3 are somewhat larger than the others because the shot noise is dominant even at relatively large scales of $\ell \gtrsim 70$ (corresponding to an angular scale of $\theta \lesssim 150$ arcmin).

Note that the radial distribution of halos is discontinuous at every $450 \, h^{-1} {\rm Mpc}$ distance, which is caused by the boundaries between the different box sizes (see Table \ref{nbody}).
As a result of this effect, we could not calculate the three-dimensional halo clustering across the boundaries.
However, note also that these discontinuities do not affect any two-dimensional observables (e.g., cosmic shear, galaxy-galaxy lensing, or CMB lensing). These will affect only three-dimensional observables like a halo power spectrum (or correlation function).

\subsection{Power Spectra of CMB Anisotropies}

\begin{figure}
\hspace*{-0.5cm}
\includegraphics[width=1.2\columnwidth]{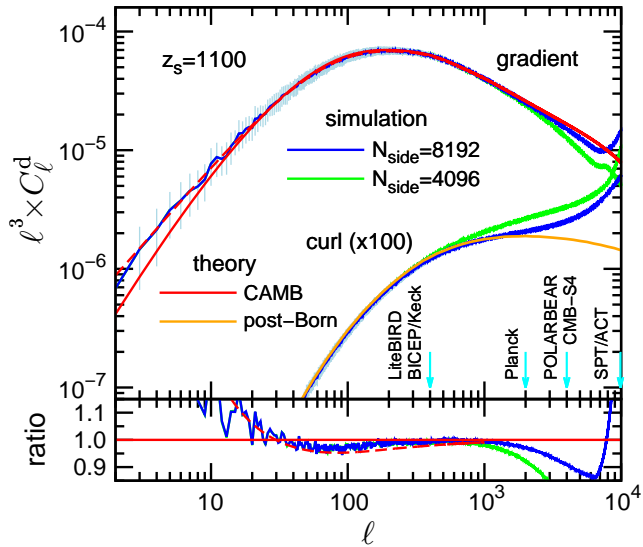} 
\caption{Deflection-angle power spectrum $\ell^3 \times C_\ell^{\rm d}$ at the last scattering surface ($z_{\rm s}=1100$).
  Here, we plot the gradient and curl modes separately (with the curl mode $C_\ell^{\rm d}$ multiplied by $100$).
  The blue and green curves show the average simulation results from the $108$ maps at $N_{\rm side}=8192$ and $4096$, respectively.
  The light-blue bars give the standard deviations. 
  The solid red (orange) curve is the theoretical prediction calculated by CAMB (using the post-Born corrections in \cite{kh10}).
  The dashed red curve is the same as the solid red curve (CAMB), but includes the lens-shell thickness effect.
  The down cyan arrows indicate the typical angular resolutions of CMB experiments.
  The bottom panel shows the ratios of the gradient mode results to those in CAMB.
}
\label{fig_cl_alpha}
\end{figure}

\begin{figure*}
  \includegraphics[width=2.\columnwidth]{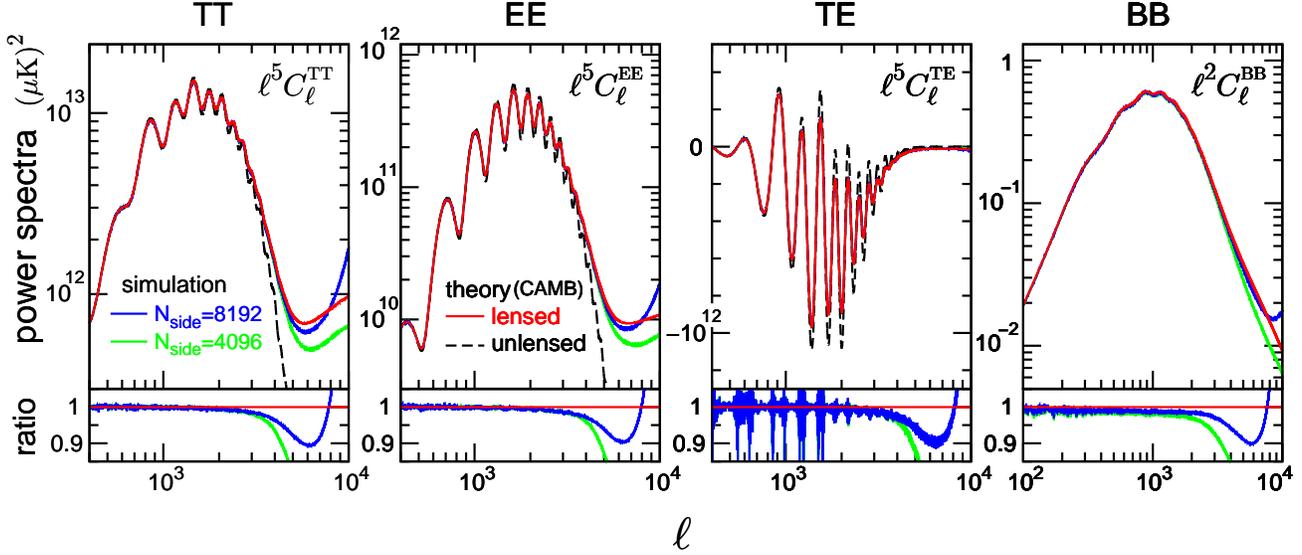} 
\vspace*{-4.3cm}
\caption{
  Angular power spectra of CMB temperature (T) and polarization (E/B-mode) anisotropies, given as TT, TE, EE, and BB from left to right.
  The $y$-axes give $\ell^5 C_\ell$ $(\ell^2 C_\ell)$ for the TT,TE, and EE (BB) to show the results clearly.
  The blue and green curves give the averages from the $108$ CMB maps.
  The solid red (dashed black) curves show the theoretical prediction of CAMB for the lensed (unlensed) case.
  The bottom panels show the ratios to the solid red curves.
}
\label{fig_cl_cmb}
\end{figure*}

\begin{figure}
\hspace*{-1.cm}
\vspace*{-0.5cm}  
\includegraphics[width=1.2\columnwidth]{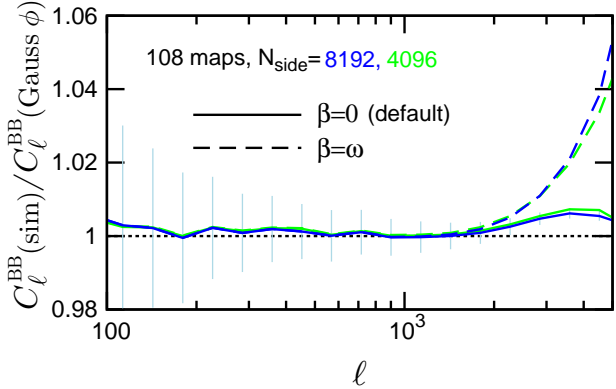} 
\caption{
  Ratio of the CMB B-mode power spectra from our simulation $C_\ell^{\rm BB}({\rm sim})$ to that based on the Gaussian lensing potential using the Born approximation $C_\ell^{\rm BB}({\rm Gauss}\,\phi)$.
  The power spectra are averages from the $108$ maps.
  The results are binned with $\Delta \log_{10} \ell=0.1$.
  The blue and green curves correspond to $N_{\rm side}=8192$ and $4096$, respectively.
  The solid curves assume $\beta=0$ in Equation (\ref{QU_trans}), while the dashed curves assume $\beta=\omega$.
}
\label{fig_cl_ratio_curl_bb}
\end{figure}

In this subsection, we calculate the angular power spectra of the CMB deflection angle and lensed CMB temperature and polarization. 
The deflection angle can be decomposed into two parts using the parity symmetry, $\bfd=\nabla\phi+(\star\nabla)\Omega$, where $\star$ is the $90^\circ$ rotation operator\footnote{ The $a$th component of the derivative is $(\star\nabla)^a = \epsilon^{ab} \nabla_b$ where $\epsilon^{ab}$ is the two-dimensional Levi-Civita symbol.}, and the first and second terms are the gradient and curl modes, respectively \citep{hs03}.
The scalar and pseudo-scalar lensing potential, $\phi$ and $\Omega$, are obtained by solving the Poisson equations $\nabla^2 \phi = \nabla \! \cdot \bfd$ and $\nabla^2 \Omega = (\star\nabla) \cdot \bfd$, respectively.
Although the gravitational lensing by density fluctuations in linear theory does not generate the curl mode under the Born approximation, the multiple-lens scattering employed in our simulation leads to a non-zero curl mode \citep{pl16}. 
This curl mode is also generated by any vector or tensor metric perturbation \citep{nyt12,saga15}. 

Figure \ref{fig_cl_alpha} shows the power spectrum of the deflection angle for both modes: the upper (lower) curves correspond to the gradient (curl) mode. 
We used the \texttt{HEALPix} subroutine (\texttt{alm2map$\_\,$spin}) to calculate the power spectrum for each mode separately.
The blue (green) curves with light-blue bars show averages with standard deviations for the $108$ maps with $N_{\rm side}=8192$ $(4096)$.
The solid red curve shows the theoretical prediction by CAMB, in which we employed the revised Halofit for the nonlinear matter power spectrum.
The orange curve shows the leading correction to the Born approximation (the so-called ``post-Born correction'') predicted by the third-order perturbation of the gravitational potential \citep{kh10}.
We also plot downward pointing cyan arrows to indicate typical angular resolutions of CMB experiments: $\ell \sim 400$ ($\theta \sim 30$ arcmin) for BICEP, Keck, and LiteBIRD; $\ell \sim 2000$ ($\theta \sim 5$ arcmin) for Planck; $\ell \sim 4000$ ($\theta \sim 3$ arcmin) for POLARBEAR and CMB-S4; $\ell \sim 10^4$ ($\theta \sim 1$ arcmin) for SPT and ACT.
The bottom panel shows the ratio of simulation results to the CAMB output.
The simulation results show an enhancement at very large scales, $\ell \lesssim 20$, and a small suppression at intermediate scales, $30 \lesssim \ell \lesssim 200$, arising from the lens-shell thickness, as also seen in Figure \ref{fig_cl_conv}.
The large-scale enhancement can reach $30\% (70\%)$ at $\ell=10 (2)$.
The discrepancies between the simulation and CAMB are well fitted by including the lens-shell thickness effect\footnote{Here, we replaced the matter power spectrum in CAMB, as in Equation (\ref{pk_replace}).} (the dashed red curve).
For smaller scales, $\ell \gtrsim 3000$, the simulation results gradually deviate from the theoretical predictions owing to the lack of the angular resolution.
For the much smaller scales of $\ell \sim 10^4$, the simulation results show an increase as a result of the shot noise.
Because the particle number density decreases at higher redshift in our simulation-box configuration (see Table \ref{nbody}), the shot noise becomes more significant at higher source redshifts.
We also numerically checked that the deflection-angle power spectrum satisfies the relations $C^{\rm d}_\ell=4 C^\kappa_\ell / [\ell (\ell+1)]$ for the gradient mode and $C^{\rm d}_\ell=4 C^\omega_\ell / [\ell (\ell+1)]$ for the curl mode (here $C^\omega_\ell$ is the angular power spectrum of the rotation, and we compare the simulation results with theoretical predictions for $C^\omega_\ell$ in Appendix C). 
The curl-mode power spectrum agrees with the leading-order post-Born correction at $\ell<1000$. 
These results indicate that as predicted by \citet{saga15}, the contributions of the vector and tensor perturbations from the nonlinear density fluctuations are negligible compared to those from the post-Born corrections at $\ell<1000$. 

We next examine the angular power spectra of CMB anisotropies.
We calculated the power spectra of the temperature (T) and polarization (E/B-mode) from the $108$ CMB maps.
Figure \ref{fig_cl_cmb} plots the results for TT, TE, EE, and BB in the left to right panels, respectively, as a function of $\ell$.
The blue and green curves show the mean values calculated from the $108$ realizations.
The solid red (black dashed) curves give the lensed (unlensed) power spectra calculated by CAMB.
The simulation results are suppressed at small scales ($\ell \gtrsim 10^3$) owing to the lack of angular resolution and rise again at very small scales ($\ell \gtrsim 5 \times 10^3$) as a result of the shot noise.
The effect of lens-shell thickness does not seem to appear in the CMB lensing because the shell-thickness influences the deflection-angle power spectrum at large scales $\ell < 200$, but the lensing effect on the CMB is not significant at such large scales.
The bottom panels show the ratios to the theoretical model (CAMB) results.
The ratio for $C_\ell^{\rm TE}$ seems noisy because $C_\ell^{\rm TE}$ crosses zero at several multipoles and the ratio diverges there.
Our simulation results agree with the theoretical prediction within $5\%$ for $\ell<4000$ $(2000)$ for $N_{\rm side}=8192$ $(4096)$.

To evaluate the contribution of the curl mode to the B-mode spectrum, we compared the B-mode spectrum derived above with results of a CMB map simulation in which the lensing potential, $\phi$, was generated as a random Gaussian field. 
Note again that the lensed CMB maps based on the Gaussian lensing potential with the Born approximation do not contain the curl mode.
We prepared the CMB maps as follows: (1) we first measured the average convergence power spectrum $C_\ell^\kappa$ at $z_{\rm s}=1100$ from the $108$ maps and obtained the angular power spectrum of the lensing potential using the relation $C_\ell^\phi=4 C_\ell^\kappa / [ \ell (\ell+1) ]^2$; (2) we generated the deflection-angle field on the sphere based on $C_\ell^\phi$ assuming Gaussian fluctuations and then created the lensed CMB maps from the unlensed maps using \texttt{LensPix} (as in Section 2.4). We prepared $108$ CMB maps in this manner.

Figure \ref{fig_cl_ratio_curl_bb} shows the ratio of the B-mode power spectrum in our simulation $C_\ell^{\rm BB}({\rm sim})$ to the spectrum based on the Gaussian lensing potential $C_\ell^{\rm BB}({\rm Gauss}\,\phi)$.
Each curve represents a mean and error calculated from the $108$ maps.
 The solid curves assume $\beta=0$ (which is our default) in Equation (\ref{QU_trans}), while the dashed curves assume $\beta=\omega$.
As shown by the solid curves, the post-Born deflection angle enhances the B-mode power spectrum by $0.5-1 \%$ at very small scales $\ell \gtrsim 2000$. 
The dashed curves show stronger enhancement than the solid curves, indicating that the polarization rotation leads to the further E-to-B leakage, which is consistent with a recent analytical calculation \citep[][Figure 2]{maro16} and a numerical ray-tracing simulation \citep[][Figure 17]{fcc17}.
The future CMB-S4 experiment will detect this tiny effect \citep{fcc17}.
In the TT, TE, and EE power spectra, the enhancements are smaller than $0.1\%$ $(1\%)$ at $\ell<5000$ for $\beta=0$ ($\beta=\omega$).
As the B-mode is fully generated by lensing (here we ignore any primordial B-modes), the curl mode is more prominent in $C_\ell^{\rm BB}$ than in the other CMB power spectra.

\section{Covariances of Observables}

In this section, we calculate the covariances of the convergence power spectrum (Section 4.1) and the CMB B-mode power spectrum (Section 4.2).
We then compare these to a simple Gaussian variance.
We also discuss variances in regions taken from the all-sky maps for the shear correlation functions and halo-galaxy lensing (Section 4.3).
We ignore the shape noise for these analyses.

\subsection{Covariance of the Convergence Power Spectrum}

\begin{figure}
\hspace*{-1.cm}
\vspace*{-0.5cm}  
\includegraphics[width=1.2\columnwidth]{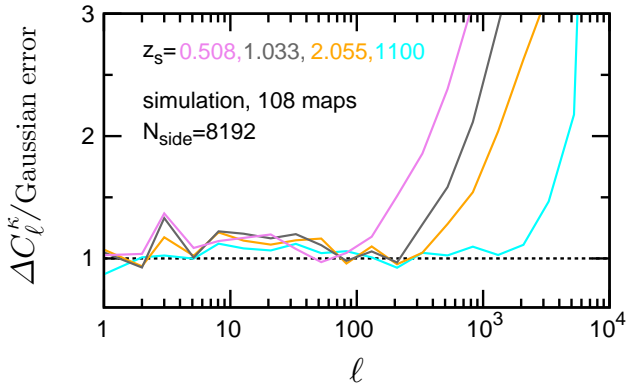} 
\caption{
  Standard deviation of $C_\ell^\kappa$ normalized by the Gaussian error for $z_{\rm s}=0.508,1.033,2.055,$ and $1100$.
  The variance is calculated from the $108$ maps with a bin width of $\Delta \log_{10} \ell = 0.2$. 
}
\label{fig_cl_conv_error}
\end{figure}

\begin{figure}
\hspace*{-1.cm}
\includegraphics[width=1.2\columnwidth]{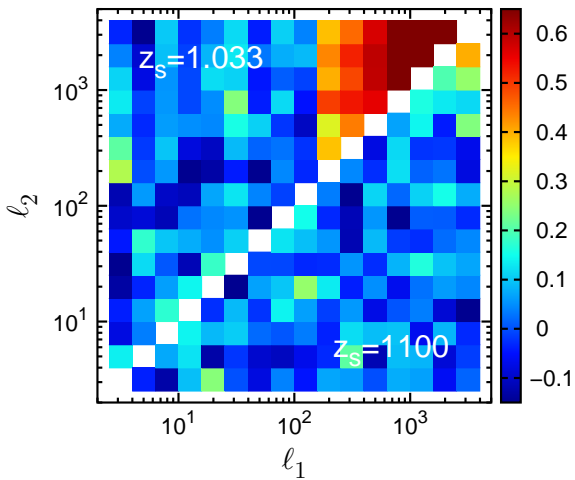} 
\caption{
  Contour plot of the correlation matrix of the convergence power spectrum calculated from the $108$ maps.
  The upper left part corresponds to $z_{\rm s}=1.033$, while the lower right part corresponds to $z_{\rm s}=1100$.
  The result is for $N_{\rm side}=8192$ with a bin width of $\Delta \log_{10} \ell = 0.2$.
}
\label{fig_cl_conv_corr}
\end{figure}

We evaluate the covariance of the convergence power spectrum $C_\ell^\kappa$ from the $108$ simulation maps with the main goal of checking the independence of the maps.
Because we constructed the $108$ maps from the $84$ $N$-body realizations, the maps are not strictly independent.
The covariance of $C_\ell^\kappa$ can be written as \citep[e.g.,][]{mw99,szh99}
\beqa
{\rm cov}(\ell_1,\ell_2) &\equiv& \langle \left( C_{\ell_1}^\kappa - \langle C_{\ell_1}^\kappa \rangle \right) \left( C_{\ell_2}^\kappa - \langle C_{\ell_2}^\kappa \rangle \right)  \rangle \nonumber \\
 &=& \frac{2}{N_{\ell_1}} \left( C_{\ell_1}^\kappa \right)^2 \delta_{\ell_1 \ell_2} + T_{\ell_1 \ell_2}.
\label{cl_var}
\eeqa
The first term gives the Gaussian variance, $N_\ell (=2 \pi \ell \Delta \ell)$ is the number of modes between $\ell$ and $\ell + \Delta \ell$, and $\delta_{\ell_1 \ell_2}$ is the Kronecker delta.
The second term, written by the trispectrum, is the non-Gaussian contribution arising from mode coupling during nonlinear gravitational evolution.

Figure \ref{fig_cl_conv_error} shows the standard deviation of $C_\ell^\kappa$ normalized by the Gaussian error (the square root of the first term in Equation (\ref{cl_var})) with the source redshifts $z_{\rm s}=0.508$, $1.033$, $2.055$, and $1100$ at $N_{\rm side}=8192$.
Here, we use the average $C_\ell^\kappa$ measured from the simulation maps to calculate the Gaussian variance.
The power spectrum is binned with the width $\Delta \log_{10} \ell = 0.2$.
The figure clearly shows that the errors are consistent with the Gaussian error for large scales ($\ell < 100$), which suggest that the maps can be treated as independent\footnote{However, the variance seems $\sim 20\%$ larger than the Gaussian prediction at $\ell=1-100$, which might be caused by the finite shell-thickness effect. This effect slightly changes the convergence power spectrum at $\ell \lesssim 100$ (see Section 3.1) and thus may also affect its covariance at such large scales.}.
The non-Gaussian error is more significant at smaller scales ($\ell > 100$), especially for lower source redshift.
Note that because only the Gaussian error depends on the bin width in Equation (\ref{cl_var}), the relative importance of the non-Gaussian term is less significant at smaller bin-widths.
Finally, we note that according to previous studies \citep[e.g.,][]{hss07,tjk13}, the typical accuracy of the variance estimated from the $108$ samples is $(108/2)^{-1/2} \sim 14 \%$.
The scatter seen in the figure at $\ell < 100$ is roughly consistent with this estimate and thus is caused by the finite number of samples.

We next examine the off-diagonal elements of the covariance matrix.
Here, we define the correlation matrix as
\beq
  {\rm corr}(\ell_1,\ell_2) = \frac{{\rm cov}(\ell_1,\ell_2)}{\sqrt{{\rm cov}(\ell_1,\ell_1)\, {\rm cov}(\ell_2,\ell_2)}}.
\eeq
The diagonal elements of this matrix are always unity by definition, while the off-diagonal elements measure the relative strength of correlation between different multiples.
Figure \ref{fig_cl_conv_corr} shows a contour plot of the correlation matrix.
The top left (bottom right) triangle corresponds to $z_{\rm s}=1.033$ $(1100)$.
As seen in the figure, the correlation is very weak at large scales ($\ell < 100$) at both redshifts; however, a strong correlation is seen at smaller scales ($\ell \gtrsim 200$) at $z_{\rm s}=1.033$.

\cite{p16a} recently demonstrated that one or two $N$-body realizations are sufficient to construct $\sim 10^4$ independent weak-lensing maps based on an investigation of the numerical convergence of the covariances of the convergence power spectrum and peak count by varying the number of $N$-body simulations.
In their Figure 3, there does not seem to be a correlation between the number of N-body simulations and the amplitude of the $C_\ell^\kappa$ variance. For instance, the smaller N-body realizations tend not to generate the smaller (or larger) variance.
According to their results, our $84$ simulations are large enough to construct independent maps.

\subsection{Covariance of the CMB B-mode Power Spectrum}

\begin{figure}
\hspace*{-1.cm}
\vspace*{-0.5cm}  
\includegraphics[width=1.2\columnwidth]{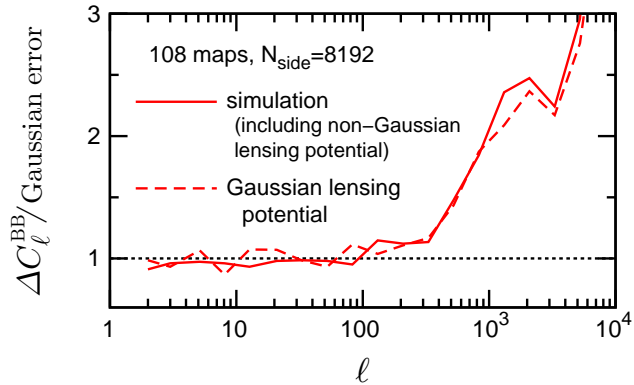} 
\vspace*{-0.5cm}
\caption{
  Standard deviation of the CMB B-mode power spectrum normalized by Gaussian error.
  The variance is calculated from the $108$ maps.
  The solid curve is calculated from the simulation maps (including the non-Gaussian lensing potential), while the dashed curve is calculated from the maps based on the Gaussian lensing potential with the Born approximation. 
  Here, we set $N_{\rm side}=8192$ with a bin width of $\Delta \log_{10} \ell = 0.2$.
}
\label{fig_cl_cmb_bb_error}
\end{figure}

\begin{figure}
\hspace*{-1.cm}
\includegraphics[width=1.2\columnwidth]{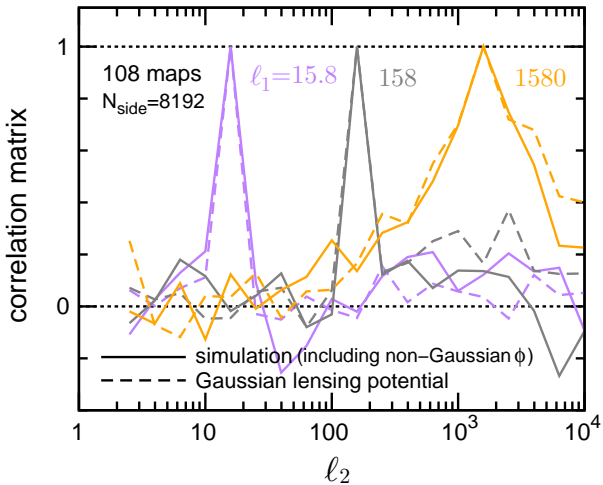} 
\caption{
  Correlation matrix of the CMB B-mode power spectrum calculated from the $108$ maps as a function of $\ell_2$ with $\ell_1=15.8$ (purple), $158$ (gray), and $1580$ (orange), respectively. 
  The solid curves are calculated from the simulation maps (including the non-Gaussian lensing potential $\phi$), while the dashed curves are from the maps based on the Gaussian lensing potential.
  The result corresponds to $N_{\rm side}=8192$ with a bin width of $\Delta \log_{10} \ell = 0.2$.
}
\label{fig_cl_cmb_bb_corr}
\end{figure}

We next calculate the covariance of the CMB B-mode power spectrum.
As the B-mode is fully generated by gravitational lensing, it contains more non-Gaussianity than either the temperature or E-mode polarization \citep[e.g.,][]{bsh12}.
This non-Gaussianity increases the sampling error in the B-mode power spectrum and makes the B-mode less sensitive to cosmology.

Figure \ref{fig_cl_cmb_bb_error} plots the standard deviation of the CMB B-mode power spectrum $C_\ell^{\rm BB}$ calculated from the $108$ maps. 
The solid curve corresponds to $C_\ell^{\rm BB}({\rm sim})$ measured in our simulation, while the dashed curve corresponds to $C_\ell^{\rm BB}({\rm Gauss}\,\phi)$ based on the Gaussian lensing potential with the Born approximation described in Section 3.5.
To calculate the Gaussian variance in the denominator, we used the average $C_\ell^{\rm BB}$
for each case.
As seen in Figure \ref{fig_cl_cmb_bb_error}, the two curves agree well up to $\ell=5000$, suggesting that the lowest-order lensing deflection is the main source of the non-Gaussianity and the multiple-scattering does not make a strong contribution. 

Figure \ref{fig_cl_cmb_bb_corr} plots the correlation matrix as a function of $\ell_2$ for various multipole values, $\ell_1=15.8$, $158$, and $1580$.
As in Figure \ref{fig_cl_cmb_bb_error}, the solid curves show $C_\ell^{\rm BB}({\rm sim})$, while the dashed curves show $C_\ell^{\rm BB}({\rm Gauss}\,\phi)$.
As there are no clear differences between the solid and dashed curves (although there are large scatters), Figures \ref{fig_cl_cmb_bb_error} and \ref{fig_cl_cmb_bb_corr} suggest that estimating the covariance of the CMB B-mode power spectrum based on the Gaussian lensing potential with the Born approximation would be valid up to $\ell \simeq 5000$.
However, we note that more realizations with higher resolutions are necessary to resolve these differences clearly.

\subsection{Variance Estimated from $48$ Regions in the All-sky Map}

\begin{figure*}
\hspace*{-0.5cm}
\includegraphics[width=1.1\columnwidth]{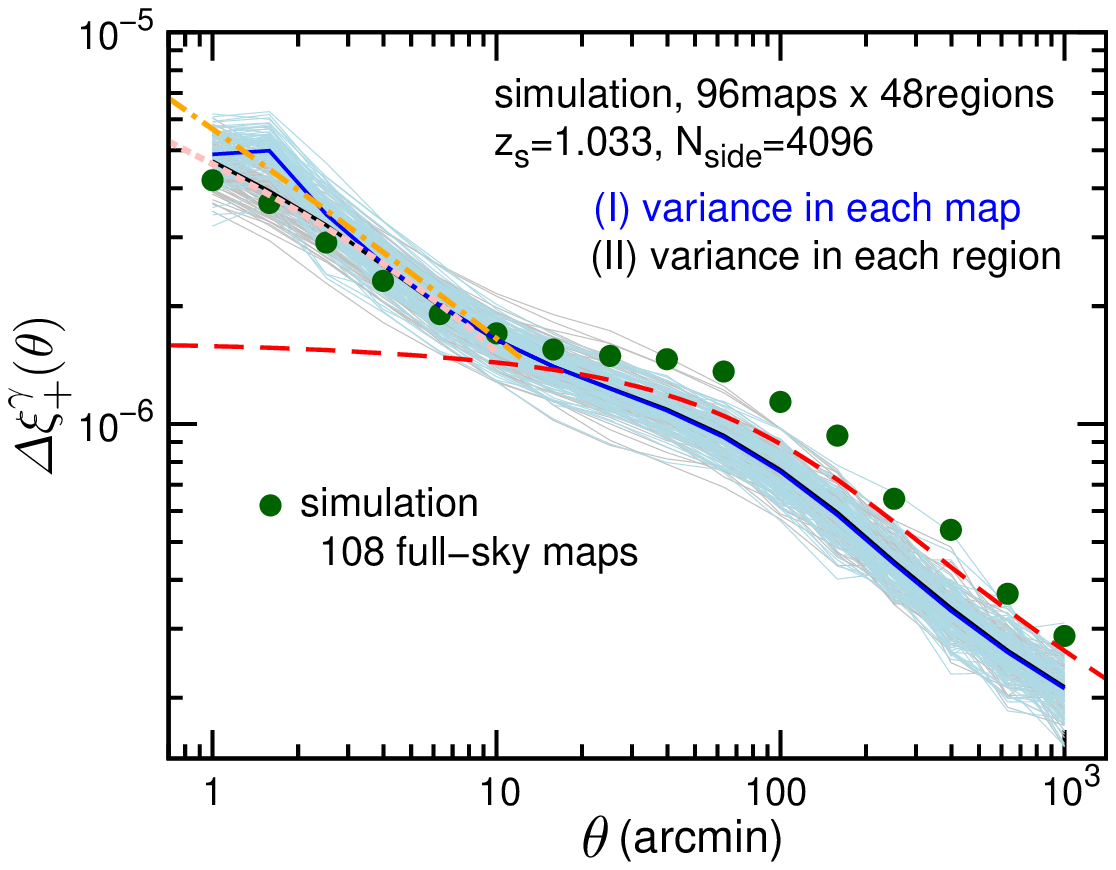} 
\hspace*{-0.5cm}
\includegraphics[width=1.1\columnwidth]{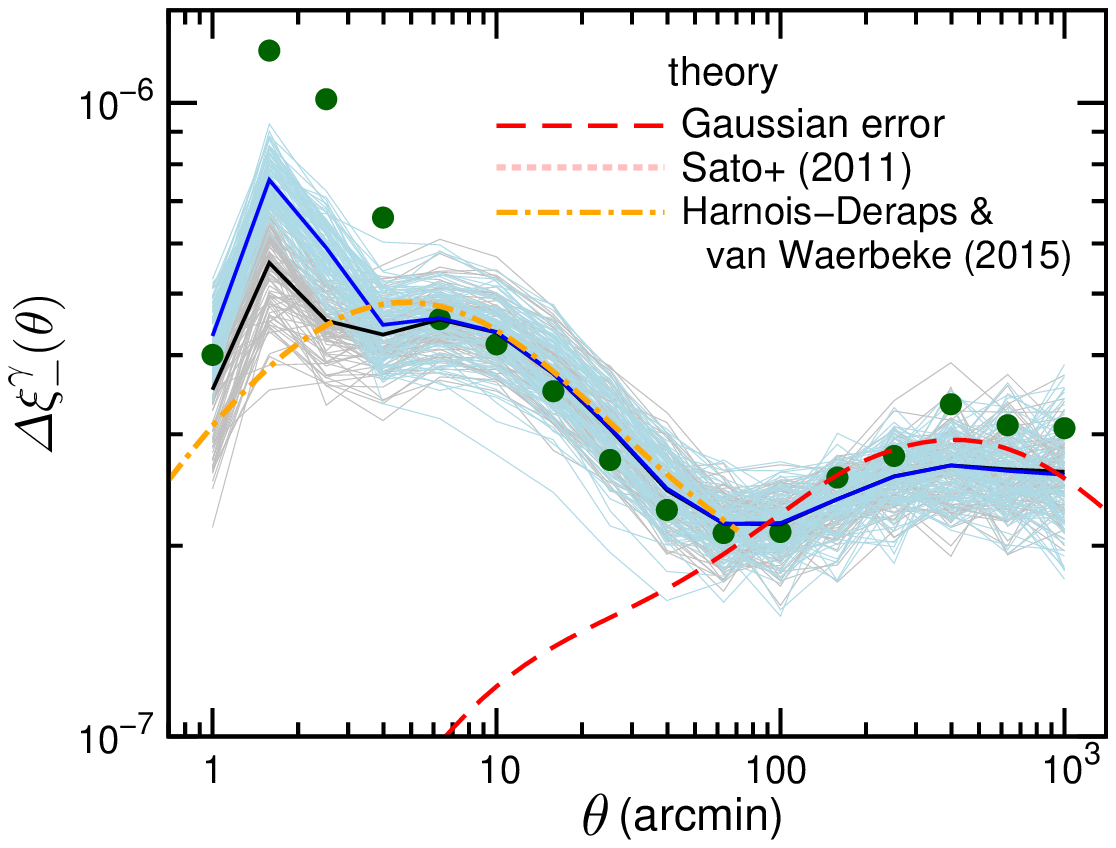} 
\caption{
  Standard deviations of the shear correlation functions at $z_{\rm s}=1.033$.
  The thin light-blue lines are the variances calculated using $48$ regions in each all-sky map (case (I)), while the thin gray lines are calculated using $48$ regions in the same sky position in the $48$ maps (case (II), see also main text).
  The thick blue and black lines show the averages of the thin light-blue and gray line sets, respectively.
  The green circles are the errors measured from the $108$ full-sky maps, multiplied by $\sqrt{48}$ to take into account the survey area ratio.
  The dashed red curves are the analytical prediction of the Gaussian error.
  The dotted pink and dash-dotted orange curves are the fitting formulae of the non-Gaussian error in \cite{sato11} and \cite{hv15}, respectively.
  The results are for $N_{\rm side}=4096$ with a bin width of $\Delta \log_{10} (\theta/{\rm arcmin}) = 0.2$.
}
\label{fig_xi_pm_error}
\end{figure*}

\begin{figure}
\hspace*{-1.cm}
\includegraphics[width=1.15\columnwidth]{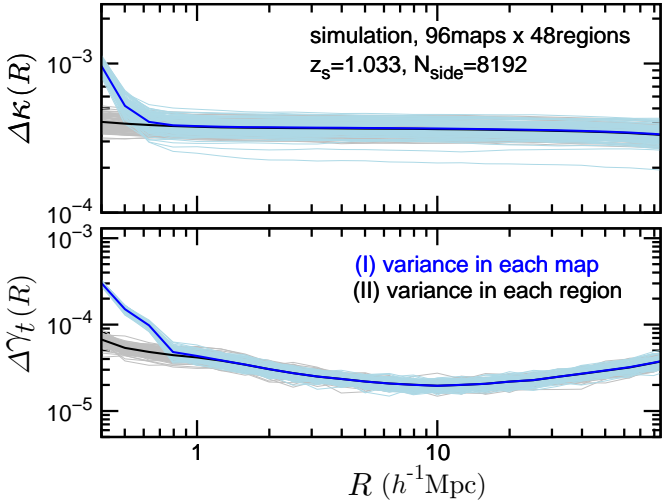} 
\caption{
  Standard deviations of the halo-galaxy lensing for sample 1 at $z_{\rm s}=1.033$.
  The results are the variances of the convergence profile (upper panel) and the tangential shear profile (lower panel).
  Here, we set $N_{\rm side}=8192$ with a bin width of $\Delta \log_{10} [R/(h^{-1}{\rm Mpc})] = 0.1$.
}
\label{fig_gglens_error}
\end{figure}

In this subsection, we assess the independence of various regions taken from a single full-sky map.
Owing to mode coupling during nonlinear gravitational evolution, density fluctuations in different survey regions correlate to each other.
As a simple example, we took $48$ equal-area regions represented by \texttt{HEALPix} large pixels with $N_{\rm side}=4$ with identical surface areas of $4\pi/48$ ($\simeq 860 \, {\rm deg}^2$).
The angular pixel resolution in each region was also the same ($N_{\rm side}=4096$ or $8192$).
We used $96$ full-sky maps to produce $4608 \, (=48 \times 96)$ regions in total and calculated the variances of shear correlation functions and halo-galaxy lensing for the two following cases.

{\it (I) Variance in each map}
We calculated the variance among $48$ regions taken from a single map.
In total, we used $96$ estimates of variance from $96$ all-sky maps.

{\it (II) Variance in each region}
We took one out of the $48$ regions from each of the $96$ maps at the same sky position.
We further divided these $96$ maps into two groups of $48$ each and computed the variance within each group to perform a straightforward comparison with case (I).
We produced $96 \, (= 2 \times 48)$ estimates of variance in total.

Figure \ref{fig_xi_pm_error} shows the standard deviations of the shear correlation functions $\xi_+^\gamma$ (left panel) and $\xi_-^\gamma$ (right panel) at $z_{\rm s}=1.033$.
We directly computed the shear correlations $\xi_\pm^\gamma$ using all pairs in the region (i.e., using the brute-force approach).
The thin light-blue and gray curves represent cases (I) and (II), respectively.
The thick black and blue curves show the averages of the thin curve sets.
The difference between the thick black and blue curves is smaller than $10 \%$ at $\theta>3(4)$ arcmin for $\xi_+^\gamma$ ($\xi_-^\gamma$). 
The blue lines are slightly larger at very small scales (close to the pixel size).
The angular resolution for $\xi_+ \, (\xi_-)$ at $N_{\rm side}=4096$ is $1 \, (10)$ arcmin (see Figure \ref{fig_xi_pm}), indicating that the discrepancies at small scale would be owing to the lack of angular resolution\footnote{Owing to pixelization, the angular separations between pixels can be sampled only discretely. The impact of this discreteness is larger on small scales close to the pixel size. This effect is common in samples in which the variance is computed in case (II) (because the pixelization in the survey region is always the same) but uncommon in case (I).
The difference in discreteness of the separation among the samples would cause differences in the variances at very small scales.}.
The green circles are the errors measured from the 108 full-sky maps. Here a factor $\sqrt{48}$ is multiplied to rescale them to take into account the differences in area. The large error seen in the right panel for $\theta<5$ arcmin would be caused by the finite angular resolution.
We also indicate the Gaussian error with the dashed red curves.
The variance of the shear correlation functions for the Gaussian fluctuations can be obtained from Equation (\ref{xi_pm_multiexp}) as \citep[e.g.,][Section VII]{nl99}
\beqa
 && \left[ \Delta \xi_{\pm ({\rm Gauss})}^\gamma(\theta) \right]^2 \equiv \langle \left[ \xi_\pm^\gamma(\theta) - \langle \xi_\pm^\gamma(\theta) \rangle \right]^2 \rangle \nonumber \\
&& ~~ = \frac{2}{\Omega_{\rm W}} \sum_{\ell=2}^{\infty} \left[ \left( C_\ell^{\gamma ({\rm E})} \right)^2 +  \left( C_\ell^{\gamma ({\rm B})} \right)^2 \right] \left[ {}_2Y_{\ell \mp 2}(\theta,0) \right]^2, \nonumber \\
\label{xi_var}
\eeqa
where $\Omega_{\rm W} (=4 \pi /48)$ is the survey area in units of steradians. 
The above equation reduces to the flat-sky result in \cite{jse08} in the small-angle limit.
In Equation (\ref{xi_var}), we adopt the spherical-sky formula for the E-mode $C_\ell^{\gamma ({\rm E})}$ and ignore the B-mode $C_\ell^{\gamma ({\rm B})}$ (as in Section 3.2).
Figure \ref{fig_xi_pm_error} shows that the blue and black curves are slightly smaller ($10-20\%$) than the Gaussian error (\ref{xi_var}) for larger scales.
\cite{sato11} showed that the theoretical Gaussian error in \cite{jse08} is less accurate and overestimates the variance for a smaller survey area of $\Omega_{\rm W} \lesssim 10^3 \, {\rm deg}^2$. The above discrepancies seem to be consistent with their finding. In other words, as they noted, the Gaussian error does not simply scale as $\Delta \xi_{\pm ({\rm Gauss})}^\gamma \propto \Omega_{\rm W}^{-1/2}$, which is also consistent with the discrepancy between the green symbols and the black (or blue) curves.
The green circles are slightly larger ($10-20 \%$) than the Gaussian error (\ref{xi_var}) for $\theta \gtrsim 100$ arcmin, which is consistent with the small enhancement of the $C_\ell^\kappa$ variance at $\ell \lesssim 100$ seen in Fig.\ref{fig_cl_conv_error}.

The non-Gaussian signature is clearly observed on the small scales, where the simulation results are well above the Gaussian estimate. We also plot the fitting formulae of the non-Gaussian error given by \cite{sato11} and \cite{hv15} \citep[see also][for an earlier fitting formula]{semb07}. They measured the covariance matrix of $\xi_\pm^\gamma$ from ensembles of weak-lensing maps based on ray-tracing simulation. According to their work, the non-Gaussian error is a product of a fitting function $F_\pm(\theta)$ and the Gaussian error defined in Equation (\ref{xi_var})
\beq
  \left[ \Delta \xi_\pm^\gamma(\theta) \right]^2 = F_\pm(\theta) \left[ \Delta \xi_{\pm ({\rm Gauss})}^\gamma(\theta) \right]^2,
\eeq
where $F_\pm(\theta)$ approaches unity for large $\theta$.  
The dotted pink and dash-dotted orange curves in Figure \ref{fig_xi_pm_error} correspond to their fitting formulae (the dotted pink curve is only for $\Delta \xi_+$).
Here we plot them only for $F_\pm(\theta)>1$.
These curves agree well with the simulation results.

Figure \ref{fig_gglens_error} shows the standard deviations for halo-galaxy lensing for sample 1 (Table \ref{halo_samples}).
The upper and lower panels show the convergence and tangential shear profiles, respectively.
Here the halos were within each region, but the background fields ($\kappa, \gamma_{1,2}$) were taken from the all-sky sample (including outside the region).
The blue and black curves agree at $R>0.6 \, (0.8) \, h^{-1} {\rm Mpc}$ for the convergence (the tangential shear) profile, but the blue curves are larger at small scales.
These features are the same as in the shear correlation function shown in Figure \ref{fig_xi_pm_error}.
The discrepancies at small scales are caused by the lack of angular resolution (which corresponds to $R \sim 1 \, h^{-1} {\rm Mpc}$ in Figure \ref{fig_gg_lens}).

In summary, we can consider the $48$ regions taken from the all-sky map to be independent samples in terms of cosmic shear and halo-galaxy lensing.
We note that the density fluctuations approach Gaussian for a higher redshift or larger survey area, in which cases the correlations between the different regions would be smaller.

\section{Known Issues}

We note that our all-sky maps have small anomalies: (i) the convergence at low source redshift ($z_{\rm s}<0.3$) can be smaller than the minimum value expected for the empty beam (which is given by setting $\delta=-1$ in Equation (\ref{eq_conv})), and; (ii) the amplitudes of the shear B-mode are slightly larger than theoretical predictions.
Feature (i) appears more frequently for lower source redshifts, and feature (ii) is probably the result of a numerical error that is caused by the finite angular resolution.
We note that these anomalies are negligibly small for most practical purposes.
A detailed discussion of the anomalies is given in Appendix C.

As shown in Section 3, the lens-shell thickness effect somewhat degrades the accuracy of our mocks. For instance, the shear correlation functions (or power spectra) decrease by at most $5\%-10\%$ at intermediate scales $\theta=10-10^3$ arcmin (or $\ell=10-100$). Similarly, for the CMB lensing, the deflection-angle power spectrum increases up to $70\%$ at very small scales $\ell=2-10$, which might be a problem for a reconstruction of the full-sky lensing potential.

The shot noise, caused by the low number density of dark matter particles, is more significant for higher source redshifts (see Table \ref{source_redshift}) and therefore especially for CMB lensing.
We note that the shot noise depends on the sky direction in our simulation setting.
The dark matter particles are placed on grids in the simulation box at the initial redshift, but do not move significantly from their initial positions at high redshifts.
Therefore, projecting the particles onto the lens shells causes them to align along the $x$-, $y$-, or $z$-axis, resulting in noisy patterns on the spheres along arcs of $\theta=\pi/2$, $\phi=0$, $\pi/2$, $\pi$, and $3\pi/2$.
This effect can be avoided if the particles are not distributed on grids in the initial redshift \citep[for example, the so-called glass initial condition; see, e.g.,][]{ww07}.
An example can be found in the middle rectangular panel of Figure \ref{fig_sample_maps}, in which a noisy pattern is apparent.
Shot noise features are significant for the convergence power spectrum, but not for the deflection angle at $z_{\rm s}=1100$ because the convergence (deflection angle) is the second (first) derivative of the lensing potential and therefore is more (less) sensitive to density contrast at the small scales close to the pixel size. 

We should also comment on the effects of baryons and neutrinos on the simulation (our simulation only includes dark matter).
Owing to baryon cooling, baryons strongly enhance matter clustering at small scales, but they suppress clustering at intermediate scales owing to feedback from active galactic nuclei (AGN), supernovae, and stellar winds \citep[e.g.,][]{vd11,vog14b}.
For instance, baryons enhance the matter power spectrum by an order of magnitude at $k \gtrsim 100 \, h {\rm Mpc}^{-1}$ but suppress it by a few tens of a percent at $k \approx 10 \, h {\rm Mpc}^{-1}$.
AGN feedback also suppresses the number density of halos by a few tens of a percent at $M \gtrsim 10^{14} \, {\rm M}_\odot$ \citep{vog14a}.
Current and future weak-lensing surveys will detect such baryonic effects via the shear correlation functions, convergence peak counts, or Minkowski functionals \citep[e.g.,][]{semb11,hvvh15,osy15}.
These baryonic effects on the matter power spectrum can be taken into account using the fitting formula given by \cite{hvvh15}.
Massive neutrinos also suppress the matter power spectrum at $k \gtrsim 0.1 \, h {\rm Mpc}^{-1}$ as a result of free streaming, and future CMB experiments will measure neutrino streaming through the CMB lensing and constrain the neutrino mass \citep[e.g.,][]{cmbs4}.


\section{Summary and Discussion}

We presented $108$ full-sky weak-lensing maps constructed using a multiple-lens ray-tracing technique through cosmological $N$-body simulations.
The full numerical simulation included nonlinear gravitational evolution, non-Gaussian error, collapsed objects (i.e., halos), and the post-Born corrections.
The resulting maps contain convergence, shear, and rotation fields at $z_{\rm s}=0.05-5.3$ and CMB temperature and polarization fields at $z_{\rm s}=1100$. 
These maps were prepared for every $150 \, h^{-1}{\rm Mpc}$ comoving distance (corresponding to a redshift interval of $\Delta z \simeq 0.05$ near $z=0$), thereby enabling the construction of a realistic shear map for an arbitrary source distribution.
The simulation has sufficient mass resolution to resolve the host halos of the SDSS CMASS galaxies and LRGs.
We demonstrated in Section 3 that the simulation results agree closely with theoretical predictions for cosmic shear, halo-galaxy lensing, halo clustering, and CMB lensing.
In Section 4 we confirmed the mutual independence of the $108$ maps, and furthermore showed that regions taken from the all-sky maps are effectively independent as well.
Thus, these mocks can be safely used to estimate the covariances of observables or test an analysis tool for real observational data.
The maps are freely available for download at the website\footnote{http://cosmo.phys.hirosaki-u.ac.jp/takahasi/allsky\_raytracing/}, where we provide a user's guide (which is also included in Appendix D).

Combining different observational probes enables breaking cosmological parameter degeneracies and therefore provides stronger constraints on the cosmological parameters.
For instance, although there are currently several distinct probes including cosmic shear, galaxy-galaxy lensing, galaxy clustering, cluster abundance, and CMB lensing, the cosmological constraints inferred from these probes are not independent, but rather are correlated because they trace the same underlying mass distribution.
In this situation, our mocks are quite useful for estimating the cross covariances between different observables in the cosmological likelihood analysis.

The $108$ full-sky maps are the largest data sets so far, but they are not large enough to estimate the covariances of observables in some cases, such as for cross-correlation analyses and higher-order (three- or four-point) correlation analyses for large survey areas. In such cases, the estimated covariance will be noisy because the number of data sets is very large.
Recently, \cite{mura17} used the $108$ catalogs to estimate the covariance of cluster-galaxy lensing for the redMaPPer clusters. They measured the jackknife covariance in each map and then calculated the average from the 108 maps to reduce the noisy feature.

Finally, we would like to note some related work that uses our mock catalogs.
\cite{stm16} recently discussed the covariances of cluster/galaxy-galaxy lensing for the SDSS LRG, CMASS, and redMaPPer clusters using our first $48$ maps and then compared the result to the jackknife covariance.
\cite{hi17} investigated the detectability of a supervoid via weak lensing of background galaxies.
The HSC science analyses are also currently using the catalogs in their likelihood analysis \citep[e.g.,][]{mand17,omh17}.
In these papers, the simulation and the observational data agree well, and they show how well our mocks serve their purposes.


\acknowledgements

We would like to thank Eric Hivon, Matthias Bartelmann, Masamune Oguri, Yuji Chinone, Joachim Harnois-D{\'e}raps, Antony Lewis, and Giovanni Marozzi for their useful comments and discussion.
We would like to thank M. R. Becker for making the source program of CALCLENS available, and the \texttt{HEALPix} team for making the \texttt{HEALPix} software publicly available.
This work was supported in part by JSPS KAKENHI Grant No. JP17H01131 (R.T.) and MEXT KAKENHI Grant Number 26400285 (T.H.), 17K05457 (T.H.) and 17K14273 (T.N.).
This work is in part supported by MEXT Grant-in-Aid for Scientific Research on Innovative Areas (Nos. 15H05887, 15H05892, and 15H05893).
M.S. is supported by Research Fellowships of the Japan Society for the Promotion of Science (JSPS) for Young Scientists.
T.N. acknowledges financial support from Japan Science and Technology Agency (JST) CREST Grant Number JPMJCR1414.
Numerical computations presented in this paper were in part carried out on Cray XC30 and on the general-purpose PC farm at Center for Computational Astrophysics, CfCA, of the National Astronomical Observatory of Japan.



\appendix


\section{Measurements of Matter Power Spectra, Halo Mass Functions, and Halo Biases}

\begin{figure*}
\hspace*{-0.5cm}
\includegraphics[width=1.2\columnwidth]{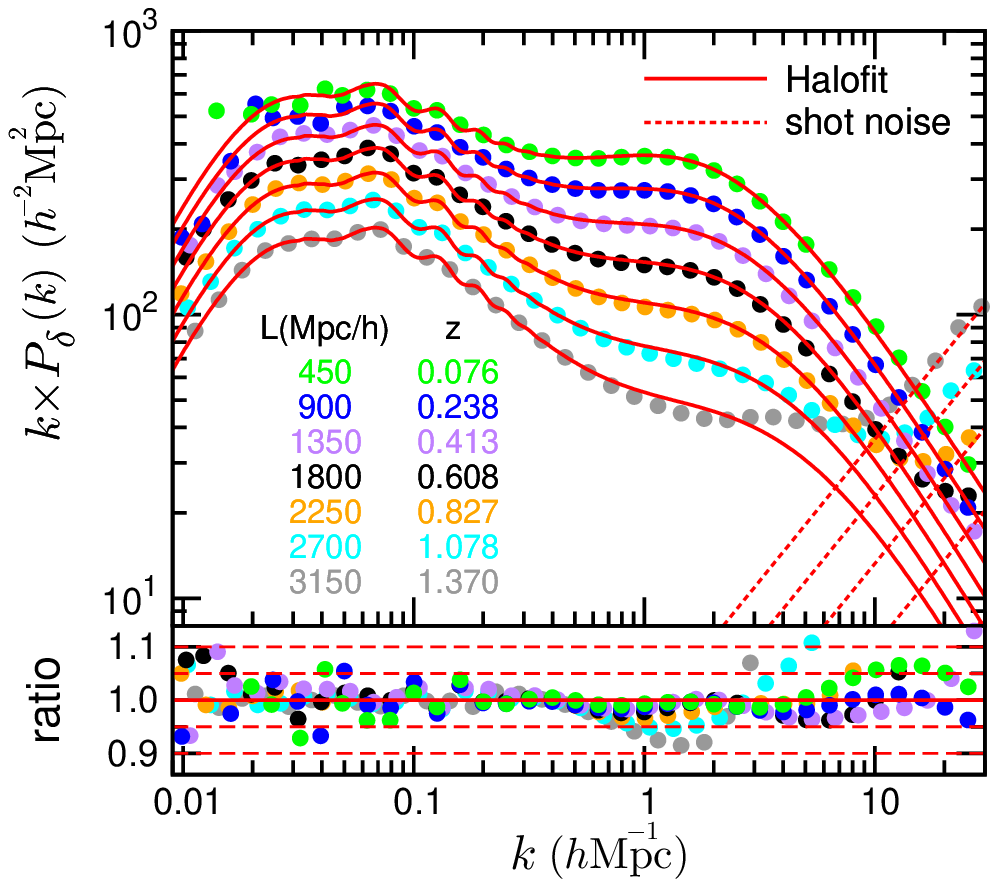} 
\hspace*{-1.2cm}
\includegraphics[width=1.2\columnwidth]{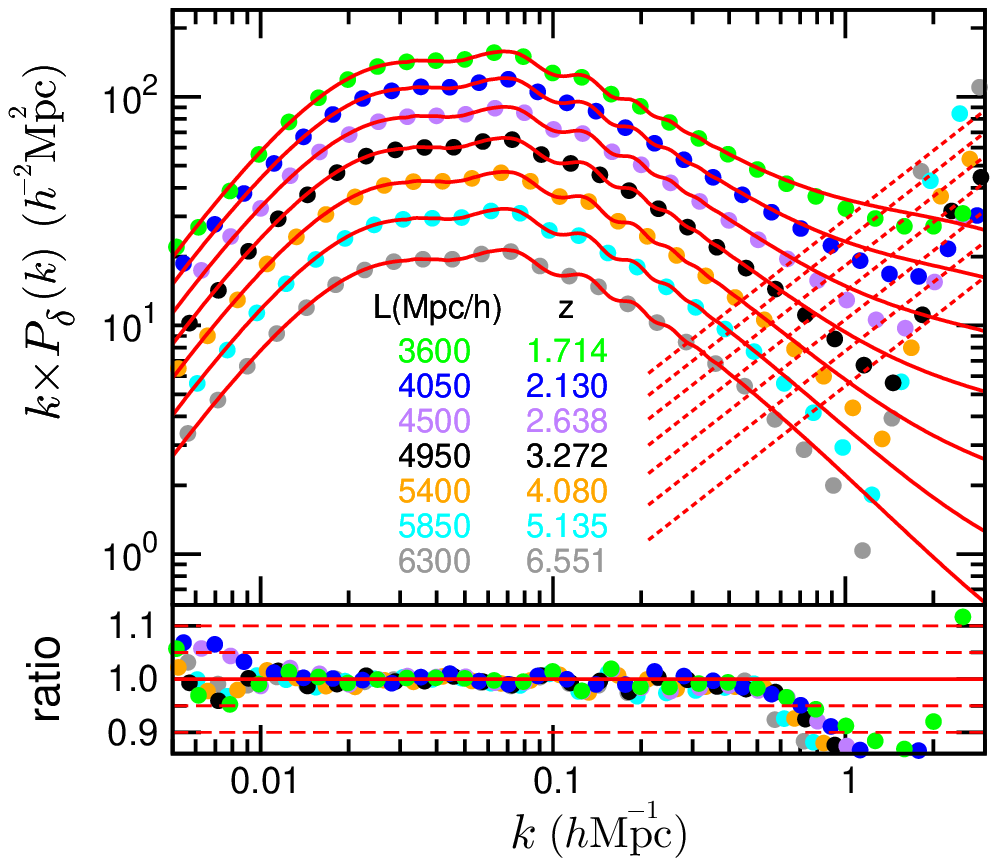} 
\caption{Matter power spectra $k \times P_\delta(k)$ as a function of wavenumber $k$ for lower redshifts (left panel) and higher redshifts (right panel).
The filled circles are the average simulation results of the six realizations with a bin width of $\Delta \log[k/(h {\rm Mpc}^{-1})]=0.1$.
Each colored symbol corresponds to a separate redshift.  
The solid red curves are the revised Halofit, and the dotted red lines are the shot noise. 
The lower small panels plot ratios to the solid red curves.
The dashed red lines denote $5 \%$ and $10 \%$ errors.   
}
\label{fig_pk}
\end{figure*}

\begin{figure}
\hspace*{-0.5cm}
\includegraphics[width=1.2\columnwidth]{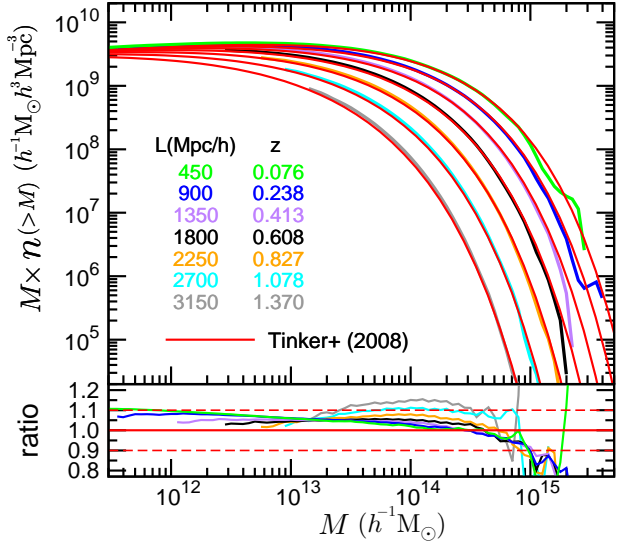} 
\caption{Halo mass functions $M \times n(> \!\! M)$ as a function of halo mass $M$ for several redshifts, where $n(> \!\! M)$ is the cumulative halo number density.
  Each colored curve corresponds to the average simulation result of six realizations at each redshift.
  The red curves are the fitting formula of \cite{t08}.
  The bottom panel plots ratios to the solid red curves.
}
\label{fig_massf}
\end{figure}

\begin{figure}
\hspace*{-0.5cm}
\includegraphics[width=1.2\columnwidth]{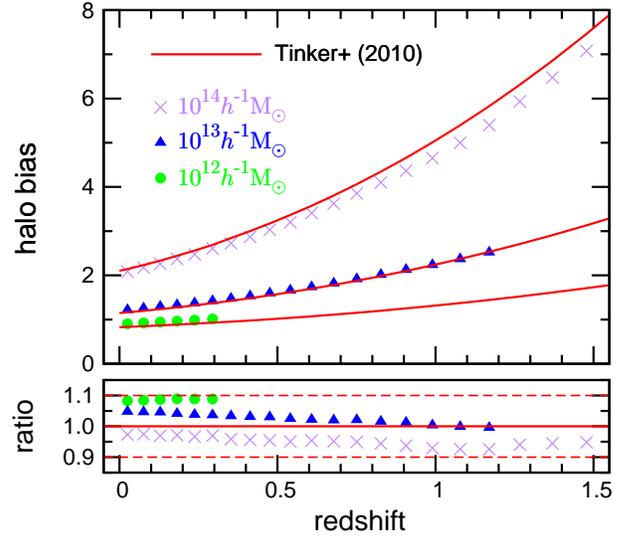} 
\caption{Linear halo biases as a function of redshift for $M=10^{14} \, h^{-1}{\rm M}_\odot$ (purple crosses), $10^{13} \, h^{-1}{\rm M}_\odot$ (blue triangles), and $10^{12} \, h^{-1}{\rm M}_\odot$ (green circles).
These colored symbols are the average simulation results of six realizations, while the red curves are the fitting formula of \cite{t10}.
The bottom panel shows ratios to the solid red curves.
}
\label{fig_bias}
\end{figure}

In this appendix, we present comparisons of our simulation results to the theoretical fitting formulae for the matter power spectra, halo mass functions and linear halo biases.

Figure \ref{fig_pk} plots the average matter power spectra measured from the six realizations for the $14$ simulation-box sizes (see Section 2.1).
Each redshift corresponds to the central dumped redshift of each box size (see Table \ref{nbody}).
Here we do not subtract the shot noise from the measured power spectrum. 
The solid red curves are the theoretical fitting formula of the revised Halofit. 
The bottom panels show ratios to the red curves.
The scatters seen at smaller $k$ $(< 0.01-0.1 \, h {\rm Mpc}^{-1})$ are caused by the sample variance.
For relatively large $k \gtrsim 1 \, h {\rm Mpc}^{-1}$, the simulation results decrease as a result of the lack of spacial resolution, which is more apparent for larger box-sizes.
For very large $k$ $(> 1-10 \, h {\rm Mpc}^{-1})$, the simulation results increase as a result of the shot noise and approach the dotted lines.
In the right panel, the dotted lines seem to fit the data poorly at intermediate scale $k \sim 1h {\rm Mpc}^{-1}$ because the simple shot-noise model (i.e., the inverse of the particle number density) overestimates the power, especially at higher redshifts \cite[see e.g.,][Sections 4 and 6.2]{emu2}.
The simulation results agree with the Halofit model within $5\%-10\%$ up to $k=3 \, h{\rm Mpc}^{-1}$ and $0.6 \, h{\rm Mpc}^{-1}$ for lower ($z \lesssim 1.5$) and higher redshifts ($z \gtrsim 1.5$), respectively.

Figure \ref{fig_massf} shows the halo mass functions measured from the six $N$-body realizations for several redshifts.
In the vertical axis, $n(>\!\!M)$ is the cumulative number density of halos heavier than $M$.
Here we adopt $M_{200 {\rm b}}$ as the halo mass $M$.
Each colored curve is the average simulation result for each redshift, which is plotted for a halo mass higher than the minimum mass (= $50 \times$ particle mass) in Table \ref{nbody}. 
The solid red curves are the theoretical fitting formula of \cite{t08}.
For much higher masses ($M \gtrsim 10^{15} \, h^{-1} {\rm M}_\odot$), the sample variance causes large scatter.
The figure shows that the measured mass functions agree with the theoretical prediction within $10\%$ up to $z=1$.

Finally, Figure \ref{fig_bias} plots the halo biases as a function of redshift for several halo masses.
Here, the masses are binned with a bin size of $0.2$ dex (i.e., $11.9 < \log[M/(h^{-1}{\rm M}_\odot)] < 12.1$ for $10^{12} \, h^{-1} {\rm M}_\odot$).
The linear halo bias is measured as the ratio of the halo power spectrum to the matter power spectrum, $b_{\rm h}^2 (M,z) = P_{\rm h}(k;M,z)/P_\delta(k;z)$, in the linear regime where the wavenumber is set to be smaller than $10$ times the fundamental mode (i.e., $k < 10 \times 2 \pi /L$).
We subtracted the shot noise from the measured halo power spectrum.
There are no data points at higher redshifts for $M=10^{12}$ or $10^{13} \, h^{-1} {\rm M}_\odot$ because $M$ is below the minimum halo mass at such high redshifts. 
The figure shows that the halo biases agree with the fitting formula of \cite{t10} within $10 \%$.

\section{Matter Power Spectrum Convolved with the Window Function of a Spherical Lens Shell}

\begin{figure}
\hspace*{-0.5cm}
\includegraphics[width=1.1\columnwidth]{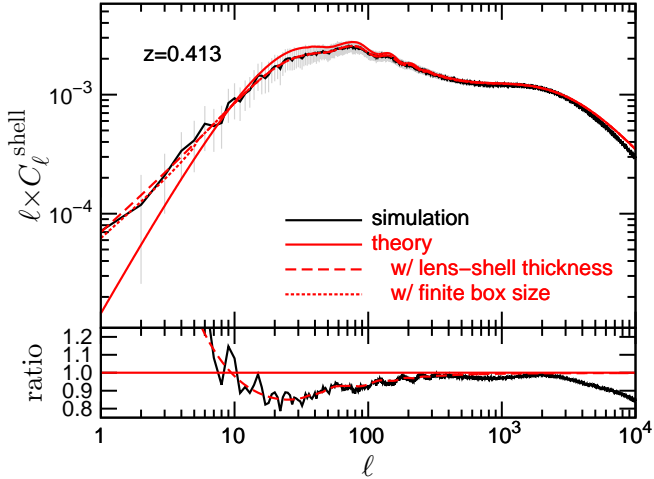} 
\caption{Angular power spectrum of the surface density on the lens shell at $z=0.413$.
  The black curve gives the average of $108$ simulation maps at $N_{\rm side}=8192$.
  The gray bars are the standard deviations.
  The dashed red curve shows the theoretical prediction including the finite thickness of the shell, while the solid red curve shows it without the correction.
  The dotted curve is the same as the dashed curve, but includes the effect of the finite-box size.
  The bottom panel shows ratios to the solid red curve.}
\label{fig_cl_shell}
\end{figure}

In this appendix, we discuss the effect of the lens-shell thickness on the angular power spectrum of density fluctuations.
We consider a spherical shell with thickness $\Delta r$ from a cubic simulation box, as shown in Figure \ref{fig_shell}.
In the box, the mean density is $\bar{\rho}$ and the density contrast is $\delta (\bfx)$.
For a shell with an inner radius $r-\Delta r/2$ and an outer radius $r+\Delta r/2$, the surface density (in units of mass per steradian) is
\beqa
\Sigma(\bftheta) &=& \bar{\rho} \, \int_{r-\Delta r/2}^{r+\Delta r/2}
 \!\! dr^\prime {r^\prime}^2 \left[ 1 + \delta(r^\prime \bftheta,r^\prime) \right],
\nonumber \\
 &=& \bar{\Sigma} \left[ 1+ \delta^{\rm shell}(\bftheta) \right],
\eeqa
where $\bar{\Sigma}=\bar{\rho} \, r^2 \Delta r$ is the mean surface density and
\beq
\delta^{\rm shell}(\bftheta) = \frac{1}{r^2 \Delta r} \int_{r-\Delta r/2}^{r+\Delta r/2} \!\! dr^\prime {r^\prime}^2 \delta(r^\prime \bftheta,r^\prime),
\eeq
is the density contrast on the shell.
Here, the density contrast $\delta$ is given as a function of the radial coordinate $r^\prime$ and the tangential coordinate $r^\prime \bftheta$, and we assume that the thickness $\Delta r$ is much smaller than the distance to the lens shell $r$.
We derive the angular power spectrum of $\delta^{\rm shell}(\bftheta)$ below.

The angular two-point correlation function of the density contrast on the shell is
\beqa
&& \xi^{\rm shell}(|\bftheta_1-\bftheta_2|) \equiv
\langle \delta^{\rm shell}(\bftheta_1) \, \delta^{\rm shell}(\bftheta_2)
\rangle  \nonumber \\
&& ~~~ = \frac{1}{\left( r^2 \Delta r \right)^2 }
 \iint_{r-\Delta r/2}^{r+\Delta r/2} \!\! dr_1 dr_2 \, r_1^2 r_2^2
 \int \! \frac{d^3k}{(2\pi)^3} \, P_\delta (k) \nonumber \\
&& ~~~ \times {\rm e}^{-{\rm i} (r_1 \bftheta_1 - r_2 \bftheta_2 ) \cdot {\bm k}_\perp}
 \, {\rm e}^{-{\rm i} (r_1 - r_2) k_\parallel},
\label{xi_shell}
\eeqa
where $k_\parallel$ and $\bfk_\perp$ are the parallel and perpendicular components of the wave vectors, respectively, and $P_\delta(k)$ is the matter power spectrum of the density fluctuation $\delta(\bfx)$ as a function of $k=(k_\parallel^2 + |\bfk_\perp|^2)^{1/2}$. 
In the first exponential, we can approximately set $(r_1 \bftheta_1 - r_2 \bftheta_2 ) \cdot \bfk_\perp \simeq r (\bftheta_1-\bftheta_2) \cdot \bfk_\perp$ with $r_1 \simeq r_2 \simeq r$.
Then, Equation (\ref{xi_shell}) reduces to
\beqa
\xi^{\rm shell}(\theta) \simeq \int \! \frac{d^3k}{(2\pi)^3} \,
P_\delta (k) \, {\rm e}^{-{\rm i} r \bftheta \cdot \bfk_\perp} \,
{\rm sinc}^2 \left( \frac{k_\parallel \Delta r}{2} \right),
\nonumber \\
\eeqa
where ${\rm sinc}(x)=\sin(x)/x$.

Therefore, the Fourier component of the correlation is
\beqa
C_\ell^{\rm shell} &=& \int \! d^2\theta \, \xi^{\rm shell}(\bftheta)
 \, {\rm e}^{{\rm i} \bftheta \cdot \bfell} \nonumber \\
 &=& \frac{1}{r^2} \int \frac{dk_\parallel}{2\pi} \, P_\delta (k)
 \, {\rm sinc}^2 \left( \frac{k_\parallel \Delta r}{2} \right),
\label{cl_shell_theory}
\eeqa
where $k=(k_\parallel^2+|\bfk_\perp|^2)^{1/2}$ with $|\bfk_\perp|=\ell/r$. 
In the limit of infinite thickness, $k_\parallel \Delta r \rightarrow \infty$, Equation (\ref{cl_shell_theory}) reduces to
\beq
\lim_{k_\parallel \Delta r \rightarrow \infty} C_\ell^{\rm shell}
= \frac{1}{r^2 \Delta r} P_\delta \! \left( k=\frac{\ell}{r} \right).
\label{cl_shell_theory2}
\eeq
Therefore, the shell thickness influences surface density fluctuations comparable to or larger than $\Delta r$, but is negligible for fluctuations much smaller than $\Delta r$.

Figure \ref{fig_cl_shell} plots the angular power spectrum of the surface density on the shell at $z=0.413$ with $\Delta r=150 \, h^{-1}$Mpc.
Here, $z \simeq 0.4$ is a typical lens redshift for the HSC survey ($z_{\rm s} \simeq 1$).
The solid black curve plots the average simulation result\footnote{Here, the simulation code \textsc{GRayTrix} adopted the cone-volume weighted distance $(3/4) [(r_2^4-r_1^4)/(r_2^3-r_1^3)]$ as the distance to the shell, which is slightly farther than the central distance of the shell $r$. This is because the volume (or the number of particles) in a farther shell with a half-radius from $r$ to $r+\Delta r/2$ is slightly larger than a closer shell from $r-\Delta r/2$ to $r$. \textsc{GRayTrix} corrected this volume difference by introducing the new distance. See also Appendix C1 of \cite{shy15}.} calculated from the $108$ maps at $N_{\rm side}=8192$. 
The dashed (solid) red curve represents the theoretical model with (without) the correction in Equations (\ref{cl_shell_theory}) and (\ref{cl_shell_theory2}).
Here, we employ the revised Halofit for the nonlinear matter power spectrum.
The theoretical model in Equation (\ref{cl_shell_theory}) agrees with the simulation results very well even for very low $\ell$ (see also the bottom panel). 
The shell thickness enhances the power at smaller scales ($\ell \lesssim 10$), but suppresses it at intermediate scales ($\ell \simeq 10-100$).
This suppression appears to be less significant in the convergence power spectrum in Figure \ref{fig_cl_conv} and in the deflection-angle power spectrum in Figure \ref{fig_cl_alpha}; this occurs because three lens shells are used as a single set in the ray-tracing simulation, resulting in an effective shell thickness of $450 \, h^{-1}$Mpc, which somewhat mitigates the thickness effect.
We confirmed that the theoretical model of Equation (\ref{cl_shell_theory}) agrees with the simulation results at other redshifts from $z=0.1$ to $7$.
However, for very low redshifts ($z < 0.1$), the simulation results show more suppression than the theoretical model of Equation (\ref{cl_shell_theory}) at intermediate scale ($\ell \approx 10$) because our assumption of $\Delta r \ll r$ does not hold for nearby lens planes, and therefore Equation (\ref{cl_shell_theory}) is less accurate.

We next comment on the effect of density fluctuations larger than the simulation-box-size $L$ on the angular power spectrum on the lens shell \citep[see also][]{sato11,hv15}.
In fact, the box does not include the fluctuations larger than $L$.
In Figure \ref{fig_cl_shell}, the dotted red curve is the same as the dashed curve, but with $P_\delta(k)=0$ imposed at $k<2\pi/L$ (here $L=1350 \, h^{-1}$Mpc) in Equation (\ref{cl_shell_theory}).
As seen in the figure, the finite-box effect is not significant compared to the lens-shell thickness effect in our setting.
The dashed and dotted curves agree completely at $\ell>6$ and within $12\%$ at $\ell \leq 6$.
The difference is much smaller than the standard deviation (the gray bars) at large scales.
The finite-box-size effect is more (less) significant at lower (higher) redshift because the box size is smaller (larger).
The finite-box-size effect is smaller than the shell-thickness effect for $L=900h^{-1} \, {\rm Mpc}$, but larger only for the second and third lens shells constructed from the $L=450 \, {\rm Mpc}$ box at $\ell<5$.
Therefore, we do not take the finite-box size into account throughout this paper\footnote{One reason that the finite-box-size effect was significant in previous work \citep{sato11} was that they adopted a smaller box size of $L=240 \, h^{-1}$Mpc.}.

In summary,  we can take the shell thickness into account by replacing the matter power spectrum as follows:
\beq
 P_\delta(k) \rightarrow P_\delta^W (k) = \Delta r \int \frac{dk_\parallel}{2\pi} \, P_\delta (k) \, {\rm sinc}^2 \left( \frac{k_\parallel \Delta r}{2} \right),
\label{pk_replace}
\eeq
from Equations (\ref{cl_shell_theory}) and (\ref{cl_shell_theory2}).
As the integration in Equation (\ref{pk_replace}) is time consuming, we construct a simple fitting function for $\Delta r = 450 \, h^{-1}$Mpc as
\beq
P_\delta^W (k) = \frac{\left( 1+ c_1 k^{-\alpha_1} \right)^{\alpha_1}}{\left( 1+ c_2 k^{-\alpha_2} \right)^{\alpha_3}} P(k),
\label{pk_replace2}
\eeq
with $c_1=9.5171 \times 10^{-4}$, $c_2=5.1543 \times 10^{-3}$, $\alpha_1=1.3063$, $\alpha_2=1.1475$, and $\alpha_3=0.62793$. The wavenumber $k$ is in units of $h {\rm Mpc}^{-1}$.
The fitting formula agrees with the correct result within $0.7\%$ at $z<7.1$.
Throughout this paper, we use the fitting function Equation (\ref{pk_replace2}) to include the shell-thickness effect.

\section{Small Anomalies in Our Simulation}

\begin{figure}
\hspace*{-0.8cm}
\includegraphics[width=1.2\columnwidth]{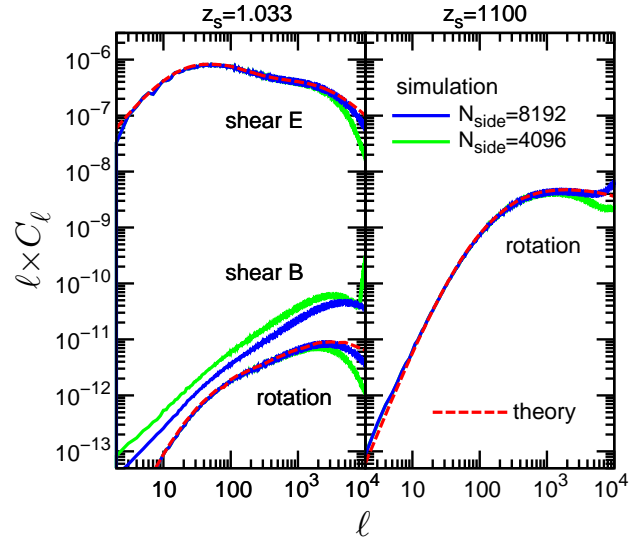} 
\caption{{\bf Left panel:} angular power spectra of the shear E-mode (upper), B-mode (middle), and rotation (lower) at $z_{\rm s}=1.033$.
  Each curve gives the average of the $108$ maps.
  The dashed red curves plot the theoretical predictions.
  {\bf Right panel:} same as the left panel, but only for the rotation at $z_{\rm s}=1100$.}
\label{fig_cl_EE_BB_cutpoles_zs16}
\end{figure}

In this appendix, we discuss the small anomalies appearing in the simulation maps, which can likely be attributed to a numerical error in our simulation.

The first anomaly is that the convergence at low redshift can be smaller than the minimum value predicted for the empty beam, which is simply obtained by setting $\delta=-1$ in Equation (\ref{eq_conv}).
The pixels with anomalous convergence are isotropically distributed over the sphere.
This anomaly appears more frequently for lower source redshifts: the fraction of anomalous-convergence pixels in the all-sky maps is $0.33 (0.22)$ at $z_{\rm s}=0.051$, $1.2 \times 10^{-2} (3.9 \times 10^{-4})$ at $z_{\rm s}=0.10$, and $1.8 \times 10^{-5} (5.2 \times 10^{-8})$ at $z_{\rm s}=0.16$ for $N_{\rm side}=8192$ $(4096)$.
Therefore, at $z_{\rm s}>0.1$, the fraction is negligibly small.
In fact, there is no anomalous convergence at $z_{\rm s}>0.3$ in the $108$ maps.
This anomaly is probably caused by a numerical error in our simulation. 

The second anomaly is an additional power seen in the shear B-mode power spectrum.
The left panel of Figure \ref{fig_cl_EE_BB_cutpoles_zs16} shows the angular power spectra of the shear E-mode (upper), B-mode (middle), and rotation (lower) at $z_{\rm s}=1.033$.
The blue (green) curves show the averages of the $108$ maps at $N_{\rm side}=8192$ $(4096)$.
The dashed red curves plot the theoretical predictions for the shear E/B-mode and rotation \citep{kh10}.
The rotation power spectrum agrees closely with the theoretical prediction that the shear B-mode and rotation power spectra should be equal.
However, the B-mode power spectrum has an additional power, which is probably caused by leakage from the E-mode to the B-mode.
The discrepancy seems more prominent for lower angular resolution ($N_{\rm side}=4096$), suggesting that the anomaly is probably caused by numerical errors related to the angular resolution in the simulation. 
We note that the B-mode power is at least three orders of magnitude smaller than the E-mode power at $\ell=2-10^4$; therefore, the additional B-mode power will be negligible for most practical purposes.

The right panel of Figure \ref{fig_cl_EE_BB_cutpoles_zs16} is the same as the left panel, but for the rotation at $z_{\rm s}=1100$.
Similar to the left panel, the rotation power spectrum agrees very well with the theoretical prediction.

\section{Guide to Using the Mock Catalogs}

Our mock catalogs can be used by visiting our website\footnote{http://cosmo.phys.hirosaki-u.ac.jp/takahasi/allsky\_raytracing/}.
It contains $108$ realizations each for $N_{\rm side}=4096$ and $8192$ and a single realization for $N_{\rm side}=16384$, which are labeled r000 to r107.
These maps with the same label (r000 $\sim$ r107) but different resolutions $N_{\rm side}$ derive from the same $N$-body data; correspondingly, they represent the same mass distribution, but simply differ in terms of angular resolution.
The lower $N_{\rm side}$ mocks are less accurate but easier to handle.
Three types of catalogs, as shown below, have been developed.

{\it Weak lensing maps.}
These contain the convergence, shear, and rotation data on the full-sky sample from source redshift $z_{\rm s}=0.05$ to $5.3$ (file ID number from zs1 to zs38) and at $z_{\rm s}=1100$ (zs66; see also Table.\ref{source_redshift}).
The data are given in the \texttt{HEALPix} pixelization with the `RING' ordering (there are $12 \times N_{\rm side}^2$ pixels on the full sky).
Each map is $3.1 \, {\rm GB} \times (N_{\rm side}/4096)^2$, as given in binary format.
We include two codes (written in C and Fortran) for reading the binary file on the site.
We also include a Fortran code to convert the binary file into the FITS format.

{\it CMB maps.}
These contain the CMB temperature and polarization (Q and U) data.
The data are given in the FITS format.
We provide the $108$ lensed CMB maps each for $N_{\rm side}=4096$ and $8192$, and also provide the $108$ unlensed CMB maps for $N_{\rm side}=8192$.
We provide the data both for $\beta=0$ and $\omega$ in Equation (\ref{QU_trans}).
The size of each map is $2.3 \, {\rm GB} \times (N_{\rm side}/4096)^2$.

{\it Halo catalogs.}
These contain the halo ID number, parent halo ID number, masses ($M_{\rm vir}, M_{\rm 200b}, M_{\rm 200c}, M_{\rm 500c}$, and $M_{\rm 2500c}$), virial radius, scale radius, redshift, comoving distance, radial peculiar velocity, angular position on the sky in both image and lens plane, and the corresponding pixel number.
The data are given in ASCII text format.
Each catalog is $40 \, {\rm GB}$ (independent of $N_{\rm side}$).
We also include the original \textsc{Rockstar} halo catalogs on the site.


\end{document}